\documentclass[11pt]{article}

\usepackage[font=small,labelfont=bf]{caption}
\usepackage{listings}
\usepackage{subcaption}
\usepackage{amsmath,amssymb}
\usepackage[left=1in, right=1in, top=1in, bottom=1in]{geometry}
\usepackage{verbatim}
\usepackage{titling}
\usepackage{parskip}
\usepackage{todonotes} 

\usepackage{graphicx} 
\usepackage[english]{babel}
\usepackage[utf8]{inputenc}
\usepackage{slashed}
\usepackage{amsfonts}
\usepackage{amsthm}
\usepackage{amsmath,amssymb}
\usepackage{empheq}
\usepackage{dsfont}
\usepackage{xcolor}
\usepackage{hyperref}

\usepackage{xcolor}

\usepackage{soul}

\title{Towards inferring network properties from epidemic data}
\author{Istvan Z. Kiss$^1$, Luc Berthouze$^2$, \& Wasiur R. KhudaBukhsh$^3$}
\date{$^1$ Department of Mathematics, University of Sussex, Falmer, Brighton BN1 9QH, UK\\
	  $^2 $ Department of Informatics, University of Sussex, Falmer, Brighton BN1 9QH, UK\\
	  $^3 $ School of Mathematical Sciences, University of Nottingham, University Park, Nottingham NG7 2RD,  UK
	  }

\begin{document}
\maketitle
\begin{abstract}
Epidemic propagation on networks represents an important departure from traditional mass-action models. However, the high-dimensionality of the exact models poses a challenge to both mathematical analysis and parameter inference. By using mean-field models, such as the pairwise model (PWM), the complexity becomes tractable. While such models have been used extensively for model analysis, there is limited work in the context of statistical inference. In this paper, we explore the extent to which the PWM with the susceptible-infected-recovered (SIR) epidemic can be used to infer disease- and network-related parameters. The widely-used MLE approach exhibits several issues pertaining to parameter unidentifiability and a lack of robustness to exact knowledge about key quantities such as population size and/or proportion of under reporting. As an alternative, we considered the recently developed dynamical survival analysis (DSA). For scenarios in which there is no model mismatch, such as when data are generated via simulations, both methods perform well despite strong dependence between parameters. However, for real-world data, such as foot-and-mouth, H1N1 and COVID19, the DSA method appears more robust to potential model mismatch and the parameter estimates appear more epidemiologically plausible. Taken together, however, our findings suggest that network-based mean-field models can be used to formulate approximate likelihoods which, coupled with an efficient inference scheme, make it possible to not only learn about the parameters of the disease dynamics but also that of the underlying network. 

\end{abstract}

Keywords: Epidemics, Networks, Inference.

\section{Introduction}

Exact mathematical models for describing the spread of epidemics on networks are often insoluble or intractable for large networks\cite{Pellis2015, kiss2017mathematics}. `Mean-field' models provide a solution by introducing approximations and focusing on quantities at the population level, such as the expectation of the number of infected or susceptible individuals, or the number of direct connections between two such groups \cite{pastor2015epidemic}. Many mean-field models exist to describe the dynamics of epidemic processes on networks. They usually take the form of a system of ODEs describing these processes
\cite{IstvanZ.Kiss2016}. Such models typically involve applying a `closure' to exact models. Closures rely on assumptions about the underlying contact network and/or even the dynamics (usually simplifying ones), and these assumptions bring the complexity of a given system to manageable levels~\cite{Sherborne2018}. 

Modelling epidemics on networks using mean-field approximations is a well studied and active area of research \cite{Porter2016,Akian2020}. In both theoretical and applied settings, it is used for parameter estimation, prediction and informing intervention or policy making \cite{Chen}, as recently demonstrated during the COVID-19 global pandemic~\cite{ReyerGerlagh2020}. However, there is a lack of understanding as to how such models operate in combination with the explicit inclusion of contact structures via networks, especially when placed in the context of statistical parameter inference. As such an investigation is warranted into whether current methods could be improved upon, or otherwise better informed, by incorporating models of epidemics on networks and by including structured population-level information and/or assumptions.

As previously mentioned, existing mean-field models are characterised by varying levels of complexity based on the assumptions used to close the exact system. This often requires making a statement about the links in the network, e.g., the number of edges that form [SI] (susceptible-infected) pairs, or [ISI] (infected-susceptible-infected) triples. For example, contact homogeneity -- that is, a fixed number of links between each node in the network -- is a common assumption~\cite{keeling1999effects,kiss2017mathematics}. In this work, we use the `pairwise' mean-field model, closed at the level of triples. Pairwise models are based on a bottom-up approach starting at node-level and building towards links and thereafter triples. This makes them  very intuitive and the `go-to choice' in many different areas. Moreover, pairwise models extend naturally to networks with heterogeneous degrees, weighted networks or even more complex epidemic dynamics. 

The aim of this paper is to investigate to what extent this model can be used for inference purposes, and more specifically, for gaining insights about both the value of the parameters of the disease dynamics and that of the contact network, thus expanding the current body of work in the field (a review of which can be found in \cite{Pare2020}). 

In Section~\ref{sec:model}, we outline the principle of epidemics on networks as stochastic processes before detailing the pairwise system of ODEs constituting the so-called mean-field SIR model. Section~\ref{sec:data}  describes simulated data -- namely, the output from the forward model with noise and Gillespie simulations, which we used to benchmark the performance of our inference schemes -- as well as three real-world datasets: (i) the 2001 UK foot-and-mouth disease outbreak, (ii) The A(H1N1) outbreak in Washington State University (WSU) campus at Pullman, and (iii) the third wave of COVID-19 in India. Section~\ref{sec:inf_methods} details the two inference schemes we considered, namely, maximum likelihood estimation and dynamical survival analysis. Section~\ref{sec:Results} presents a comparative analysis of these two schemes, both when ground-truth data is available (simulated data) and when it is not (real-world datasets). An interpretation of these results is provided in Section~\ref{sec:discussion}, along with potential new research directions.

\section{Model}
\label{sec:model}
\subsection{Epidemics on networks as a stochastic process}\label{subsec:epi_on_net_as_stoch_process}

The starting point is the modelling of population contact structures as a network of nodes connected by links
which represent possible routes of disease transmission. The network can be represented by an adjacency matrix
$G=(g_{ij})_{i,j=1,2,\dots,N}$, where $N$ is the number of nodes and the entries are either zero or one and the matrix 
is symmetric and all elements on the main diagonal are zero, i.e., no self-loops are allowed. In this paper, we will focus on
regular or homogeneous networks where each node has exactly $n$ links. 

When modelled as a continuous-time Markov Chain, a stochastic susceptible-infected-recovered (SIR) epidemic on a network results in a state space of size $3^N$ since each of the $N$ nodes can be independently 
S, I or R, and each state, that is, a labelled network, needs an equation~\cite{IstvanZ.Kiss2016}. This of course makes the model intractable both theoretically and numerically, even at modest values of $N$. Of course, Gillespie~\cite{Gillespie1976} simulations can help deal with the problem and enable us to produce true stochastic paths of the process, see Figure~\ref{fig:sim_versus_PW} for example. This is based on the simple principle
that in the Markovian framework, infection and recovery are independent Poisson process with rate $\tau$ and $\gamma$. $\tau$ is the per-link rate of infection and is the rate at which the I (infected) node in an I-S link infects the S (susceptible) node. This process is network-dependent. All infected nodes recover independently of the network and of each other at rate $\gamma$.

One way to move beyond simulations while dealing with the challenges of intractable high-dimensional models is to use mean-field models that focus on some expected quantity from the exact system, such as the expected number of infected nodes or the expected number of pairs of various types (e.g., S-S and S-I). One widely used model is the pairwise model~\cite{IstvanZ.Kiss2016} which is 
briefly described below.

\subsection{Pairwise model as an approximation of epidemics on networks}
\label{subsec:PW_as_approx}

\begin{figure}[h!]
\center
\includegraphics[scale=0.33]{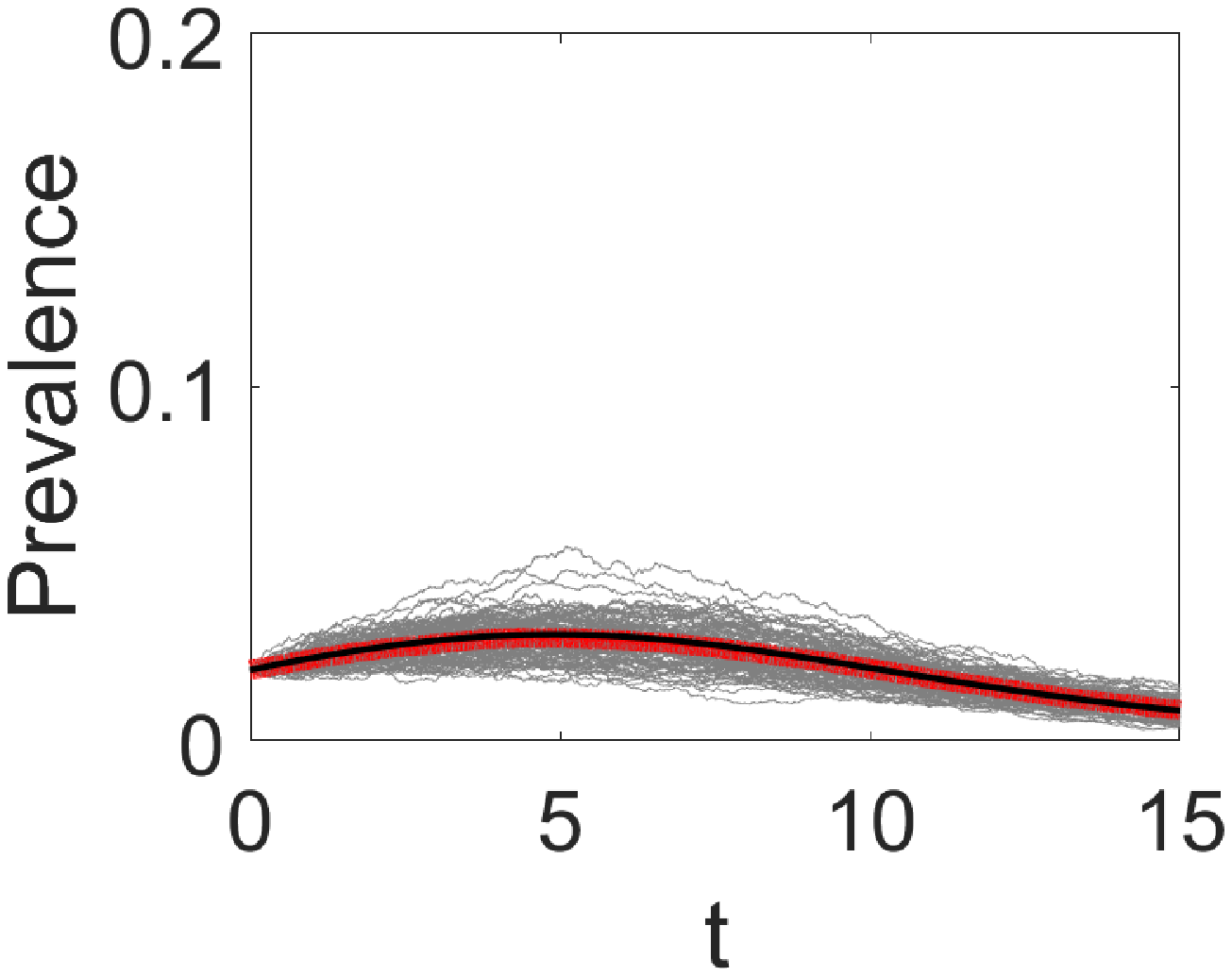}
\includegraphics[scale=0.33]{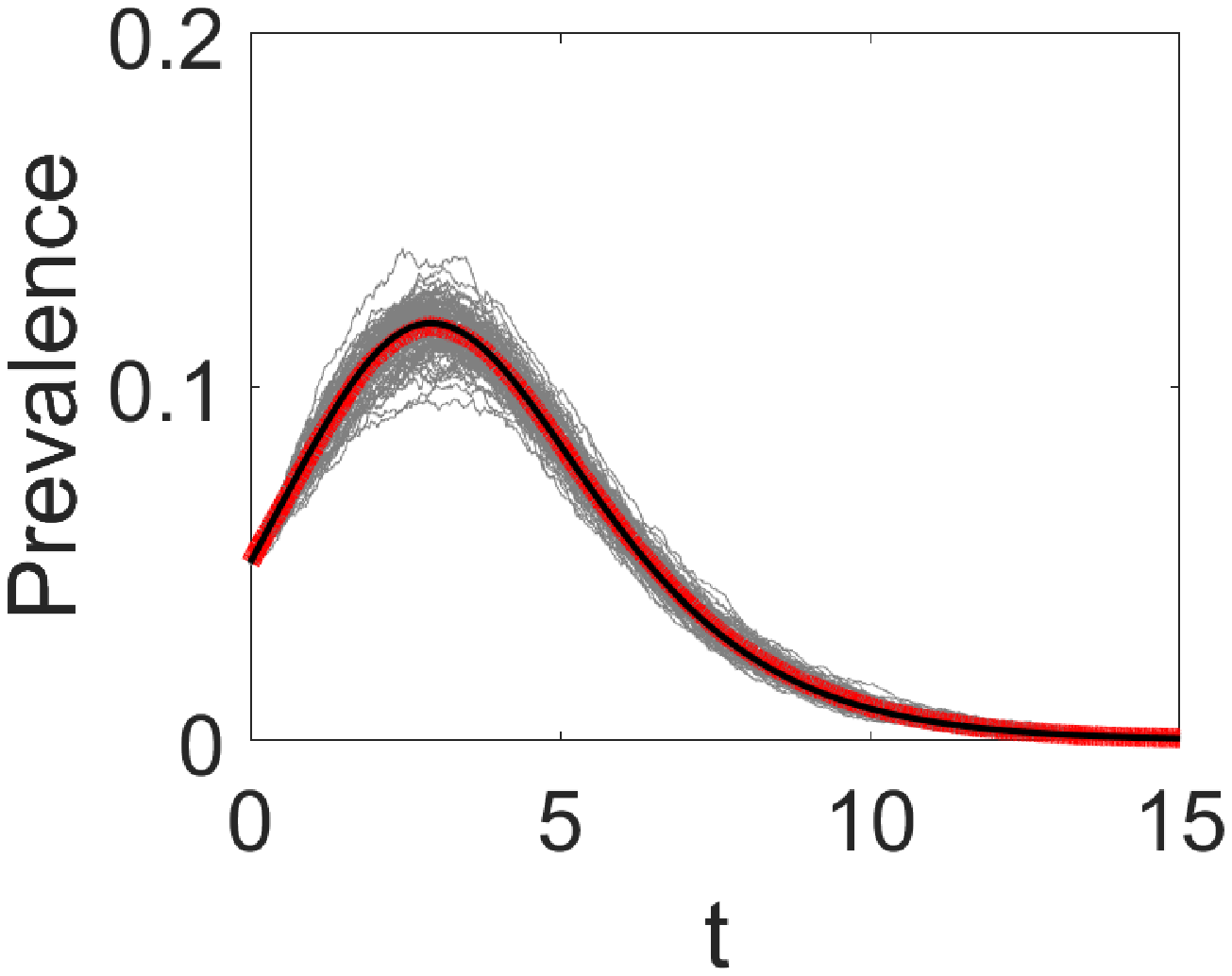}
\includegraphics[scale=0.33]{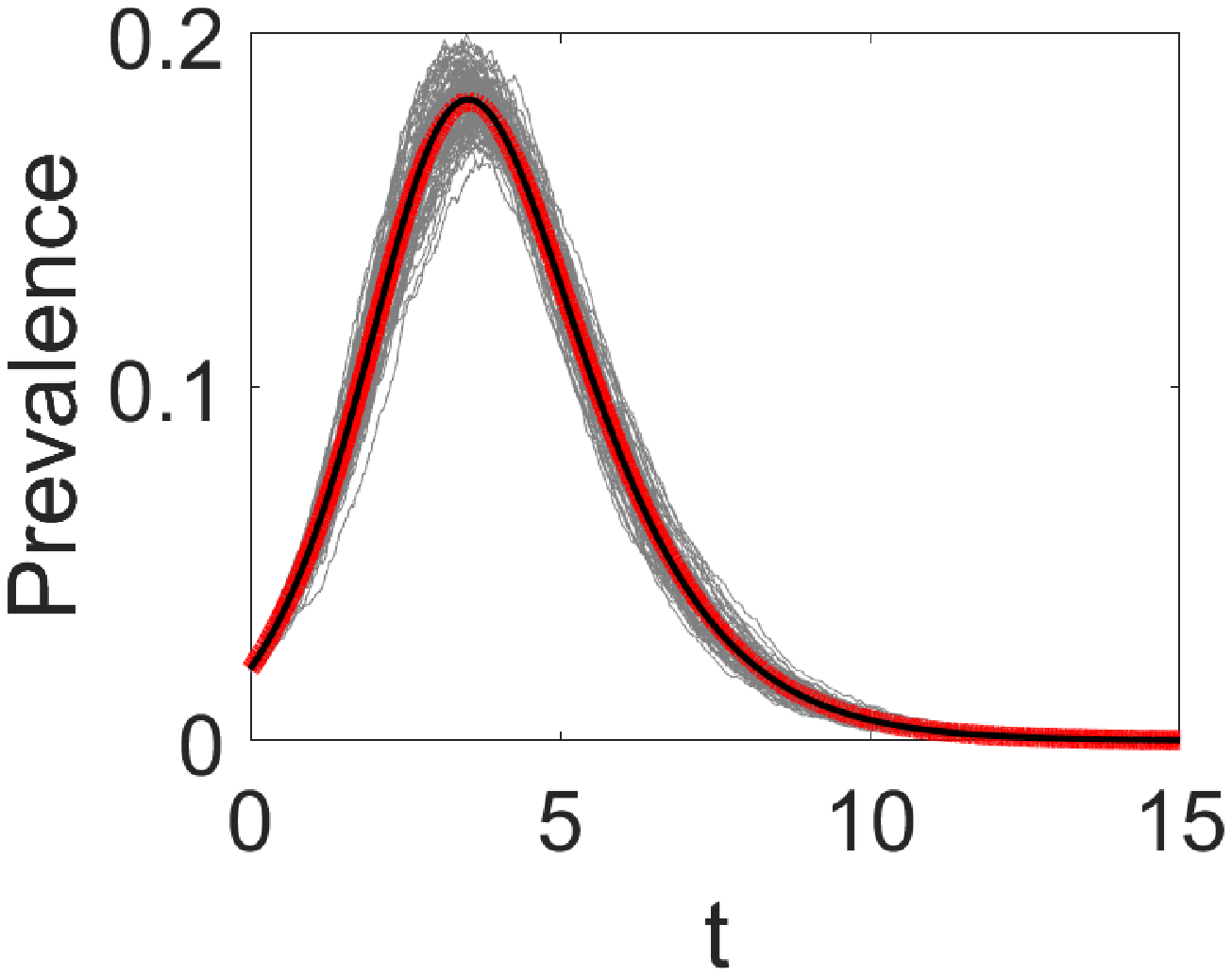}
\caption{Prevalence based on Gillespie simulations. Thin lines/cloud in grey are the outcome of $\sim 100$ individual realisations (10 networks with 10 realisations each) of an SIR stochastic epidemic on regular networks ($n=6$), with their average plotted in thick red lines. Epidemics are started with $I_0=100$ (left panel) and $I_0=250$ infectious nodes chosen at random (middle and right panels) and only epidemics that reach $2I_0$ are kept and averaged over. The numerical solution of the corresponding pairwise model is plotted as a continuous black line. All networks have $N=10000$ nodes and the recovery rate is $\gamma=1$. From left to right, $\tau$ takes value 0.3, 0.4 and 0.5, respectively.}
\label{fig:sim_versus_PW}
\end{figure}

In essence, the pairwise model focuses on a hierarchical construction where the expected number of nodes in state $A$ at time $t$, $[A](t)$, 
depends on the expected number of pairs of various types (e.g., $[AB]$) and these, in turn, depend on triples such as $[ABC]$.
Here, the counting is done in all possible directions, meaning that $[SS]$ pairs are counted twice and $[SI]=[IS]$. With this in mind, the 
pairwise model becomes
\begin{align}
[\dot{S}] &= -\tau[SI]; \,\,\,[\dot{I}] = \tau[SI]-\gamma[I]; \,\,\, [\dot{R}] = \gamma[I],\label{eq:exact_PW_singles} \\
[\dot{SI}] &= -(\tau+\gamma)[SI] + \tau([SSI]-[ISI]); \,\,\, [\dot{SS}] = -2\tau[SSI].\label{eq:exact_PW_pairs}
\end{align}
This system is not self-consistent as pairs depend on triples and equations for these are needed. This, however, would lead to an explosion in system size as triples will then depend on quadruples connected 
in ways different from the simple line graphs over four nodes. To tackle this dependency on higher-order moments, the
triples in equation~\eqref{eq:exact_PW_pairs} are closed using the following relation,
\begin{equation}
[ASB]=\frac{n-1}{n}\frac{[AS][SB]}{[S]},
\end{equation}
where $A, B \in \{A, B\}$.
Applying this closure leads to
\begin{align}
[\dot{S}] &= -\tau[SI], \\
[\dot{I}] &= \tau[SI]-\gamma[I], \\
[\dot{R}] &= \gamma[I], \\
[\dot{SI}] &= -(\tau+\gamma)[SI] + \tau\frac{n-1}{n}\frac{[SI]([SS]-[SI])}{[S]}, \\
[\dot{SS}] &= -2\tau\frac{n-1}{n}\frac{[SS][SI]}{[S]},
\end{align}
which is now a self-contained system. For a chosen set of parameters ($n, \tau, \gamma$) and initial conditions, the system above can be numerically integrated, furnishing us with $[I](t)$ for example. As it turns out, see Figure~\eqref{fig:sim_versus_PW}, this low-dimensional mean-field model is exact in the asymptotic limit of $N\rightarrow \infty$, and the numerical solution of the PW model is indistinguishable from the average of stochastic realisations. We note that there are necessary and sufficient conditions which guarantee that the PW model is exact in the limit of large network sizes. In particular, it is true for networks with Binomial (with Regular being a special case of Binomial), Poisson and Negative Binomial degree distributions~\cite{kiss2022necessary,khudabukhsh2022FCLT}.
Using that $R_0=\frac{\tau(n-1)}{\tau+\gamma}$, the closed pairwise equations can be re-parameterised to include $R_0$ explicitly. Using $\xi$ to denote $\xi=\frac{n-1}{n}$, the re-parameterised system now reads
\begin{align}
[\dot{S}] &= -\frac{\gamma R_0}{(n-1)-R_0}[SI], \label{eq:pairwise_reparam_S}\\
[\dot{I}] &= +\frac{\gamma R_0}{(n-1)-R_0}[SI]-\gamma[I], \label{eq:pairwise_reparam_I}\\
[\dot{R}] &= +\gamma[I], \label{eq:pairwise_reparam_R}\\
[\dot{SI}] &= -\left(\frac{\gamma R_0}{(n-1)-R_0}+\gamma\right)[SI] + \xi \frac{\gamma R_0}{(n-1)-R_0}\frac{[SI]([SS]-[SI])}{[S]},\label{eq:pairwise_reparam_SI} \\
[\dot{SS}] &= -2\xi \frac{\gamma R_0}{(n-1)-R_0}\frac{[SS][SI]}{[S]}.\label{eq:pairwise_reparam_SS}
\end{align}

\section{Data}
\label{sec:data}
Typically, real-world data for inference comes as daily counts of some quantity of interest (e.g., daily new cases or daily deaths) at discrete time steps, that is
\begin{equation} \label{eqn:data_prev}
\left(\mathbf{y}, \mathbf{t}\right) = \{(y_{1}, t_{1}),...,(y_{n_{obs}}, t_{n_{obs}}\},
\end{equation}
where $(y_{1},...,y_{n}) \in \{0,...N\}$ and $(t_{1},...,t_{n_{obs}}) \in \{0,T\}$ with $(0\leq t_{1}<t_{2}<\cdots< t_{n_{obs}} \leq T)$ are the counts and times respectively. 
In this paper, we will consider three types of data, which are described below.

\subsection{Data: PWM output with noise}
\label{subsec:data_PW_with_Noise}
Since the mean-field model is an approximation of the true stochastic process, we start by simulating data directly from the mean-field model and with varying levels of noise dispersion added in order to assess the ability of the inference schemes to recover the expected parameters, i.e., those used to generate the data (before noise). Since we mainly fit to daily reported cases, we first solve the PW model numerically with a given set of parameters and compute the daily new cases on day $i$, ($[S](i+1)-[S](i)$). Observations begin on the first day, at the earliest, and the initial conditions of the PWM are set at $t=0$. Noise is introduced using draws from the Negative Binomial distribution. This is done such that the mean of the distribution is given by the model and the variance is controlled by the experimenter. For the Negative Binomial, and given a daily new cases count, $y_{d}$, from the true model without noise, we draw a sample from 
\begin{equation}
X \sim NB\left(m(k)=\frac{1}{k}, p=\frac{1}{1+ky_{d}}\right),
\label{eq:negbin_random_var_dist}
\end{equation}
where the mean of this distribution is $y_{d}$, the variance is given by $y_{d}+y_{d}^2k$ with $k$ the dispersion parameter, and the negative binomial distribution is interpreted as giving the probability of observing $y_{D}$ failures given $m$ successes, that is
\begin{equation}
\mathcal{P}(X=y_{d})= {y_{d}+m-1 \choose y_{d}}p^{m}(1-p)^{y_{d}}.
\end{equation}

\subsection{Data: stochastic simulations}
\label{subsec:data_Gil_sims}
Since the real challenge is to fit to stochastic data, in the first instance, we consider simulated data constructed by using the Gillespie algorithm \cite{Gillespie1976} for a stochastic SIR epidemic  on an explicit network of contacts. The idea behind the simulation is rather simple. Each node has its own rate, resulting in a rate vector $(r_i)_{i=1,2,\dots, N}$. A susceptible node with $m$ infected
neighbours will have rate $\tau m$ and an infected node will have rate $\gamma$. Recovered or removed nodes have rate zero as they no
longer play a role in the dynamics. The time to next event is chosen from an exponential with rate $R=\sum_{i}r_i$, and the event itself will be chosen at random from all possible $N$-events but proportionally to the values of the rate, e.g., event $j$ will be chosen with probability $r_j/R$. Typical simulation plots are shown in Figures~\eqref{fig:sim_versus_PW}.

\subsection{Data: real epidemic data}
\label{subsec:data_real}

In addition to assessing the robustness of the inference schemes on synthetic data for which ground truth is known, we considered real-world outbreak data from three different data sets: 
\begin{enumerate}
    \item \emph{The 2001 Foot-and-mouth (FMD) disease outbreak in the UK}. The 2001 FMD outbreak in the UK started towards the end of February in 2001 and ended in September 2001, impacting more than 2000 farms. Control efforts resulted in the culling of millions of livestock~\cite{davies2002foot}, see Figure~\ref{fig:MLE_vs_DSA_fits}. 
    \item \emph{The A(H1N1) outbreak in Washington State University (WSU) campus at Pullman, Washington}. In April 2009, there was an outbreak of influenza virus in Veracruz, Mexico. After this initial outbreak, a new strain of the virus, A(H1N1)pdm09, started to spread around the world in the autumn. See \cite{schwartz2014estimating,KhudaBukhsh2020DSA} for more details about this triple reassortment virus, which spread even among young, healthy adults. As a result, multiple outbreaks on college campuses were seen, one of which was on the Washington State University (WSU) campus in Pullman, Washington in late August 2009. Within the space of three months, almost 2300 students came to the campus health centre with influenza-like illnesses that were treated as influenza A(H1N1) infections. Figure~\ref{fig:MLE_vs_DSA_fits} shows the daily new cases starting on 22 August 2009. 
    \item \emph{The third wave of COVID-19 in India} The COVID-19 pandemic has killed millions of people across the globe. Here, we consider the third wave in India. Similar to the other two datasets, we have daily incidence and prevalence of cases, recoveries and deaths from 15 February 2021 to 31 June 2021 (see Figure~\ref{fig:MLE_vs_DSA_fits}).
\end{enumerate}

\section{Inference methods}
\label{sec:inf_methods}

While most inference methods are based on the optimisation of a likelihood function, the likelihood function itself can be formulated based on different considerations of the underlying model and data. The most direct method typically focuses on matching model output and data as closely as possible, i.e., it is an error minimisation process. More sophisticated methods consider the underlying stochastic model in a more direct way and involve the timing of events, even if simplifying assumptions may be needed. To ensure that investigation into the possibility of inferring epidemic and network parameters using the pairwise model is not affected or biased by the inference scheme used, we consider two different methods as described below. 

\subsection{Maximum-likelihood-based approach}
In order to fit data produced by the PW model with the likelihood based on the PW model, we simply test how well the true parameters can be recovered. This scenario does not require any approximation. When fitting to stochastic data from an exact epidemic or a real epidemic, however, we are making the assumption that the exact forward model can be approximated by the PW model. 

In this paper, we use the negative-binomial distribution as likelihood of choice, because of its flexibility. The distribution models the number of failures given a target number of successes, $n$, and the probability of each experiment's success $p$. For these parameters, we have the expressions:
\begin{equation}
m(k) = \frac{1}{k},\quad  p(y(\theta,t_{i}), k) = \frac{1}{1+k y(\theta,t_{i})},
\end{equation}
with $k>0$ being the dispersion parameter, which we also attempt to infer. In this case, the distribution has mean $y_{\theta}(t_{i})$ and variance $y_{\theta}(t_{i})+y_{\theta}(t_{i})^{2}k$. This yields the following likelihood
\begin{equation}
\mathcal{L}_{NB}((\theta,k)|(\mathbf{y}, \mathbf{t})) =\prod_{i=0}^{N} {y_{i}+m-1 \choose y_{i}}p^{m}(1-p)^{y_{i}},
\label{eq:mle_likelihood}
\end{equation}

Using $\mathcal{L}_{NB}$ effectively decouples the mean and the variance of the distribution describing the data. This is expected to be sufficient to capture the variability of the data resulting from either natural stochasticity or variability due to how data was collected. 


Parameter estimation was performed by minimising the negative log-likelihood using the widely used direct search Nelder–Mead method. Because this technique can converge to non-stationary points, for each estimation process, multiple initial conditions (15) were used. To avoid biasing the search, initial conditions were drawn using Latin hypercube sampling, maximising the minimum distance between points. Because Latin hypercube sampling cannot prevent inappropriate parameter settings, initial conditions were only accepted if the ratio $\tau / \gamma$ was not too large. Specifically, we enforced that the denominator in the expression of $\tau$, i.e., $n-1-R_0$, was greater or equal than 1.5 (chosen empirically). On average, 10 out of 15 initial conditions survived. Code and data used to produce the results in Sections~\ref{sec:Results} are available from \url{https://github.com/berthouz/EpiPWMInf} for fitting the ODE realisations with negative binomial noise, \url{https://github.com/berthouz/EpiPWMInfwI0} for fitting the Gillespie stochastic realisations and  \url{https://github.com/berthouz/EpiPWMInfwI0N} for fitting the real-world datasets. The key difference between these will be explained in the relevant results sections.  

\subsection{Dynamical Survival Analysis}
\label{sec:dsa}
The statistical methodology Dynamical survival analysis (DSA) has recently been developed in a series of papers \cite{KhudaBukhsh2020DSA,DiLauro2022NonMarkovDSA,Vossler2022Ebola,khuda_bukhsh_2022_projecting} to address some of the shortcomings of traditional inference methods used in infectious diseases epidemiology. In essence, the method combines classical dynamical systems theory with tools from survival analysis. The crux of the methodology lies in interpreting the mean-field ODEs (representing population proportions) as describing  probability distributions of transfer times, such as time to infection, time to recovery. Such a change in perspective allows one to use population-level mean-field ODEs to describe the dynamics of scaled compartment sizes as well as to write a likelihood function for individual-level trajectories based on transfer times, which may be censored, truncated or even aggregated. 

To apply the DSA methodology, let us first define $[D] = [SI]/[S]$, which satisfies 
\begin{align*}
    \dot{[D]} = \tau(1-\xi)[D]^2 + \left(\xi n \tau [S]^{(2\xi-1)} - \tau -\gamma \right) [D], 
\end{align*}
with initial condition $[D](0) = n \rho$ and $[S](0) =1$, where, as before, $\xi = (n-1)/n$ and $[S]$ satisfies the pairwise mean-field equation with $[S](0)=1$ and $[I](0)=\rho$. The reason we normalize the system so that $[S](0)=1$ will be clear when we describe the DSA likelihood. Now, dividing the above equation by $\dot{[S]} = -\tau [S][D]$, solving for $[D]$ in terms of $[S]$ with initial condition $[S](0) = 1$ and then putting the solution back in $\dot{[S]} = -\tau [S][D]$, we get 
\begin{align*}
    -\dot{[S]} = n\tau\left(1- [S]^\xi\right)[S]^\xi + \frac{\gamma + \tau}{1- \xi } [S] \left(1-[S]^{\xi-1}\right) + n\tau \rho [S]^\xi,
\end{align*}
with initial condition $[S](0) =1$. In essence, DSA interprets the susceptible curve as an improper survival function for the time to infection of a randomly chosen initially susceptible individual. That is, $[S](t) = \mathsf{P}(T_I > t)$, where the random variable  $T_I$ describes the time to infection. Because $[S](t)$ is interpreted as a survival function, we set $[S](0)=1$. This survival function is improper because $\lim_{t\to \infty}[S](t) = \mathsf{P}(T_I = \infty) >0$. However, we can transform it into a proper survival function by conditioning it on a final observation time $T \in (0, \infty)$. We define the probability density function $f_{T}$ on $[0, T]$ as follows:
\begin{align}
    h_T(t) &{}= - \frac{[\dot{S}](t)}{(1- [S](T))}. 
    \nonumber
\end{align}
Given a random sample of infection times $t_1, t_2, \ldots, t_n$, the likelihood contribution of the infection times is given by 
\begin{align}
	\ell_{I}(\xi, \tau, \gamma, \rho \mid t_1, t_2, \ldots, t_n) = \prod_{i=1}^{n} h_{T}(t_i).
    \label{eq:dsa_infection_likelihood}
\end{align}
Note that DSA does not require knowledge of removal times. However, if individual recovery or removal times are known, they may be used to enhance the quality of inference. The likelihood contribution of a random sample of individual recovery times $t'_1, t'_2, \ldots, t'_m$ is given by 
\begin{align}
    \ell_{R}(\xi, \tau, \gamma, \rho \mid t'_1, t'_2, \ldots, t'_m)  &{} = \prod_{i=1}^{m} r_{T}(t'_i),
    \label{eq:dsa_recovery_likelihood}
\end{align}
where 
\begin{align*}
    r_{T}(t) &{} = \frac{ \int_0^{t} h_{T}(u) \gamma e^{-\gamma (t-u)}\mathrm{d}u}{ \int_0^T \int_0^{t} h_{T}(u) \gamma e^{-\gamma (t-u)}\mathrm{d}u \mathrm{d}t }
\end{align*}
is the density of the individual recovery times. The density $r_{T}$ is a convolution of two densities: $h_{T}$ for the time of infection and the density of an exponential distribution with rate $\gamma$ corresponding to the infectious period. In practice, it is convenient to differentiate the density $r_{T}(t)$ with respect to $t$ and then solve a system of ODEs. 

Finally, the DSA likelihood function based on a random sample of infection times $t_1, t_2, \ldots, t_n$ and a random sample of recovery times $t'_1, t'_2, \ldots, t'_m$ is given by 
\begin{align}
    \ell(\xi, \tau, \gamma, \rho \mid t_1, t_2, \ldots, t_n; t'_1, t'_2, \ldots, t'_m) = \ell_{I}(\xi, \tau, \gamma, \rho \mid t_1, t_2, \ldots, t_n)  \ell_{R}(\xi, \tau, \gamma, \rho \mid t'_1, t'_2, \ldots, t'_m). 
    \label{eq:dsa_likelihood}
\end{align}
For practical convenience (and as with the MLE-based approach), we work with the loglikelihood function, i.e., the logarithm of the likelihood function, rather than the likelihood function. It is, of course, possible to maximise the DSA likelihood function $\ell$ in equation~\eqref{eq:dsa_likelihood} to get point estimates of the parameter set $(\xi, \tau, \gamma, \rho)$. Such a procedure would then be called a maximum likelihood approach and the difference between the two inference schemes discussed here would simply be that they maximise two different likelihood functions. An alternative way to perform parameter inference using DSA is to adopt a semi-Bayesian approach via a Laplace approximation to the posterior. In this paper, we adopted a fully Bayesian approach. Specifically, we drew posterior samples of $(\xi, \tau, \gamma, \rho)$ using a Hamiltonian Monte Carlo (HMC) scheme implemented in the \emph{Stan} programming language \cite{rstan} interfaced with \textbf{\textsf{R}}. The code will be made available upon request. 

Some of the datasets used in this paper (see relevant sections) provides daily new infection cases, rather than infection and/or recovery times. As mentioned earlier, the DSA methodology does not require knowledge of removal times. When these are not available, one can simply work with the likelihood function $\ell_{I}$ (or the corresponding loglikelihood) in equation~\eqref{eq:dsa_infection_likelihood}. Infection times, in turn, can be constructed from daily new cases as follows: If we observe 10 new cases on day $t$, then we simply draw 10 random samples from a uniform distribution over $[t-0.5, t+0.5]$. By repeating this procedure for all days for which daily new case counts are available and combining the individual infection times (samples from the uniform distributions), we can transform the original count data into data on infection times. A random sample of those infection times can then be fed into the likelihood function $\ell_{I}$ in equation~\eqref{eq:dsa_infection_likelihood}. 
In datasets in which daily recoveries are available, we can construct individual recovery times in a similar fashion: If we observe 5 recoveries on day $t$, we draw a random sample of size 5 from a uniform distribution over $[t-0.5, t+0.5]$. We repeat this procedure for all days for which we have daily number of recoveries available, and then combine the individual recovery times. A random sample of this data on individual recovery times is then fed into the likelihood function $\ell_{R}$ in equation~\eqref{eq:dsa_recovery_likelihood}. 

\section{Results}
\label{sec:Results}

\subsection{ML-based inference using data produced by the PW model}\label{sec::PWMInf}
As a very first step toward assessing the ability of the inference scheme to recover the expected parameters, we first fitted the PW model (see Eqs.~\eqref{eq:pairwise_reparam_S}-\ref{eq:pairwise_reparam_SS}) to daily cases data generated by the PW model and contaminated by some noise, whose dispersion was manipulated as will be described. Here, initial conditions for parameters $R_0$, $k$, $n$ and $\gamma$ were taken from $[0.2,10]$, $[0.00001, 0.05]$, $[3, 20]$ and $[0.001, 0.1]$ respectively. 

The top row of Figure~\ref{fig:expected_param_full_epidemic} shows the histograms of parameters obtained when fitting $M=1000$ data-series, i.e. solving Eqs.~\eqref{eq:pairwise_reparam_S}-\ref{eq:pairwise_reparam_SS} with true $[R_0, n, \gamma, I_0] = [2, 6, 1/14, 1]$ and $N=10000$. Here, noise was simulated according to Eq.~\eqref{eq:negbin_random_var_dist} using $k = 0.0005$ (i.e., very low dispersion). These results confirm that the mean values are close to the true parameters, which is expected because the value of $k$ is very small.

To illustrate the sensitivity of the estimation process to the value of the dispersion parameter, we repeated the fitting process when considering 5 levels of dispersion, from $0.0005$ to $0.01$. As shown by the bottom left panel in Figure~\ref{fig:expected_param_full_epidemic}, as the dispersion level increases, so does the range of inferred $R_0$ values. Nevertheless, the mean estimated value remains close to the true value in all cases. 

Likewise, we found the inference process to be robust to the choice of time horizon (full epidemic $t_{max}=150$, partial epidemic including the peak $t_{max}=80$, epidemic up to the peak $t_{max}=70$, partial epidemic not including peak $t_{max}=60$). As shown by the bottom right panel in Figure~\ref{fig:expected_param_full_epidemic}, as the time horizon reduces, the range of inferred $R_0$ values increases but the average remains close to the true value. Importantly, whilst the inclusion of the peak does narrow the range of inferred values, it is not necessary for the inference process to correctly recover the expected value of $R_0$. 

\begin{figure}[h!]
	\center
	\includegraphics[scale=0.4]{ODEfits.png}
	\caption{Inferring $[R_0, n, \gamma, k]$ based on $M=10^3$ data realisations generated using $ [R_0, n, \gamma, k] = [2, 6, 1/14, 5\times 10^{-4}]$ with $N=10^{4}$, $I_0=1$. Dashed lines indicate the true values of the parameters. Expected values were $[1.999, 6.005, 0.0714, 0.00061]$, respectively.}
	\label{fig:expected_param_full_epidemic}
\end{figure}

\subsection{Identifiability}\label{sec::Identif}

\begin{figure}[h!]
\center
\includegraphics[scale=0.40]{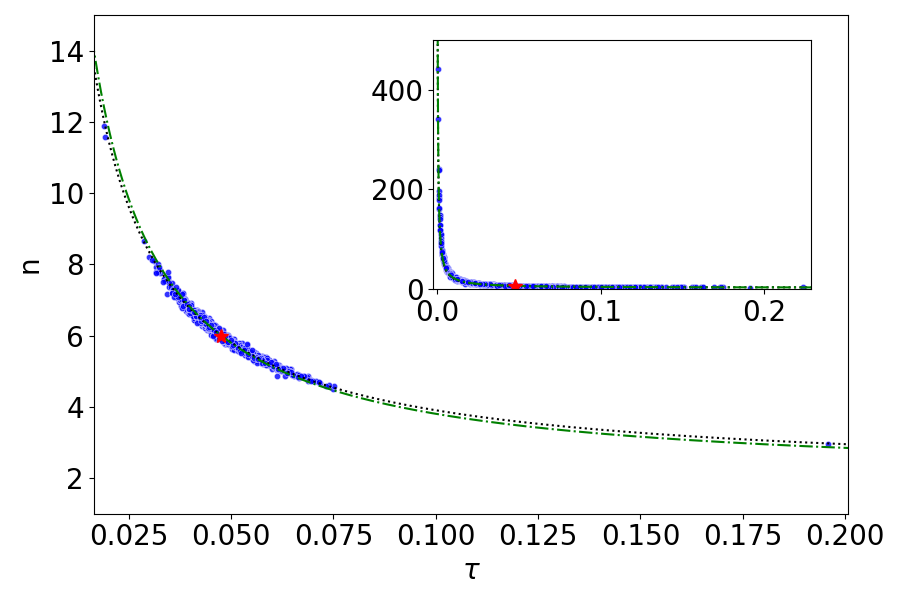}
\caption{Scatter plots of the parameter estimates on the $n$, $\tau$ plane with the two unidentifiability curves calculated as per Eqs.~\ref{eq:tau19} (dotted line), and ~\ref{eq:tau20} (dashed line). The star denotes the true values, i.e., true $n$ and calculated value of $\tau$ given true values of $R_0$ and $n$. Main panel: scatter plot when the full epidemic is used for inference. Inset: scatter plot when the time horizon does not include the peak, i.e., $t_{max}=60$. Note that an arbitrary cut-off of $n<500$ was used for clarity of the plot.}  
\label{fig:ui_curve_horizon150}
\end{figure}

As Fig.~\ref{fig:ui_curve_horizon150} shows, the inferred values of $\tau$ and $n$ describe a hyperbola-like curve which indicates a clear identifiability problem; that is the values of $\tau$ and $n$ cannot be disentangled. However, we make two important remarks. First, it is possible to characterise this hyperbola analytically. Second, the values of $\tau$ and $n$ combine favourably into the expression of $R_0$ whose inferred values are well behaved, see bottom panels in Fig.~\ref{fig:expected_param_full_epidemic}. 

To formally characterise the hyperbola, we rely on quantities that can be derived analytically from the PW model. These are the leading eigenvalue (or growth rate under some transformation) and the final epidemic size. These are given below in terms of $\tau$ as a function of $n$.
\begin{align}
\tau&=\frac{\lambda_{L}^*+\gamma^*}{n-2}, \label{eq:tau19}\\ 
 \tau&= \gamma \frac{{s_{\infty}^{*}}^{1/n}-{s_{\infty}^{*}}^{2/n}}{{s_{\infty}^{*}}^{2/n} - s_{\infty}^{*}} , \label{eq:tau20}
\end{align}
where $\lambda_L^*$ and $s_{\infty}^{*}=S_{\infty}^{*}/N$ are obtained by setting all parameters to some desired values, $(n, \tau, \gamma)=(n^*, \tau^*, \gamma^*)$; note that often $R_0$ instead of $\tau$ is given, with knowing the value of either being sufficient to have a well-defined system. The growth rate follows from the linear stability analysis of the pairwise model at the disease-free equilibrium, while the implicit formula for the final epidemic size can be found in~\cite{kiss2017mathematics} and is used here with initial conditions corresponding to the disease-free steady state.

\subsection{ML-based inference using data from exact stochastic simulations}\label{sec::Gillespie}

\begin{figure}[h!]
	\center
	\includegraphics[scale=0.45]{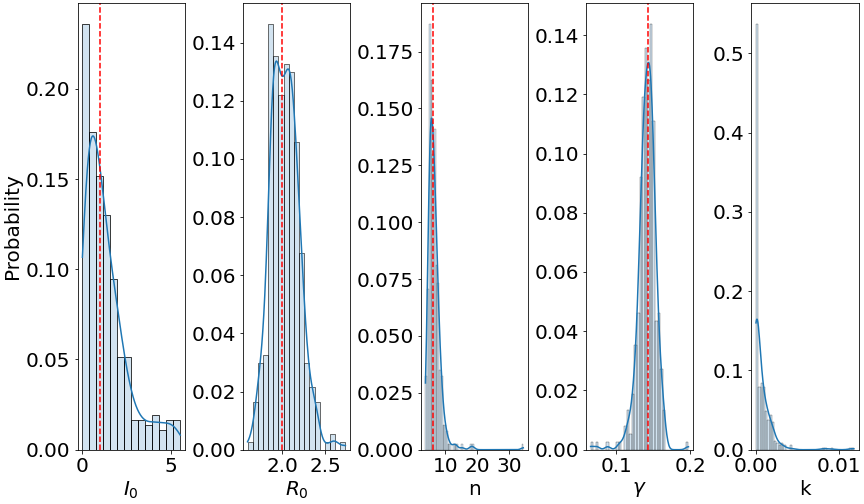}
	\caption{Inferring $[I_0, R_0, n, \gamma, k]$ based on $M=370$ data realisations generated using $ [I_0, R_0, n, \gamma] = [1, 2, 6, 1/7]$ with $N=10^{4}$. Dashed lines indicate the true values of the parameters. Mean estimated values were $[1.355, 2.029, 6.522, 0.141, 0.0071]$, respectively.}
	\label{fig:expected_param_gillespie}
\end{figure}

Five hundred Gillespie realisations were generated using parameters $[R_0, n, \gamma, I_0] = [2, 6, 1/7, 1]$ and $N=10000$. Of these $500$ realisations, $M=370$ realisations did not die out. Figure~\ref{fig:expected_param_gillespie} shows the histograms of the parameters estimated from fitting those realisations. Unlike with noisy realisations of the ODE, we also subjected $I_0$ to the inference process. Results (not shown) obtained when assuming $I_0=1$ during estimation revealed that the inclusion of $I_0$ in the estimation process was key to being able to account for the stochasticity in the onset of the epidemic, or more precisely, the time elapsed before the growth becomes exponential. For the purpose of initialising Latin hypercube sampling, values were taken in $[0.01, 10]$. This particular choice has no bearing on our findings (results not shown). The mean of the estimated $I_0$ was found to be $1.355$, i.e., close to the expected $1$; however, it showed a broad distribution of values, ranging from $0.012$ to $5.534$. 

Comparing these histograms to those shown in Figure~\ref{fig:expected_param_full_epidemic}, we find that whilst the mean estimated values do not significantly differ, the variance in estimation is, not surprisingly, substantially larger. To quantify this more precisely, we calculated the mean (and standard deviation) of the confidence intervals on $R_0$ over all $M=370$ realisations. Specifically, we determined the nominal $99\%$ profile likelihood confidence interval widths for $R_0$ as described in ~\cite{King2015}. Confidence intervals are $0.534 \pm 0.203$ compared to $0.498 \pm 0.071$ when fitting the ODE realisations with noise (dispersion level $k=0.0005$). These results are representative of those obtained when calculating confidence intervals for the other parameters (not shown).

\subsection{Inference based on DSA}
\label{sec:DSA_principle}

\begin{figure}[h!]
    \centering
    \includegraphics[scale=0.325]{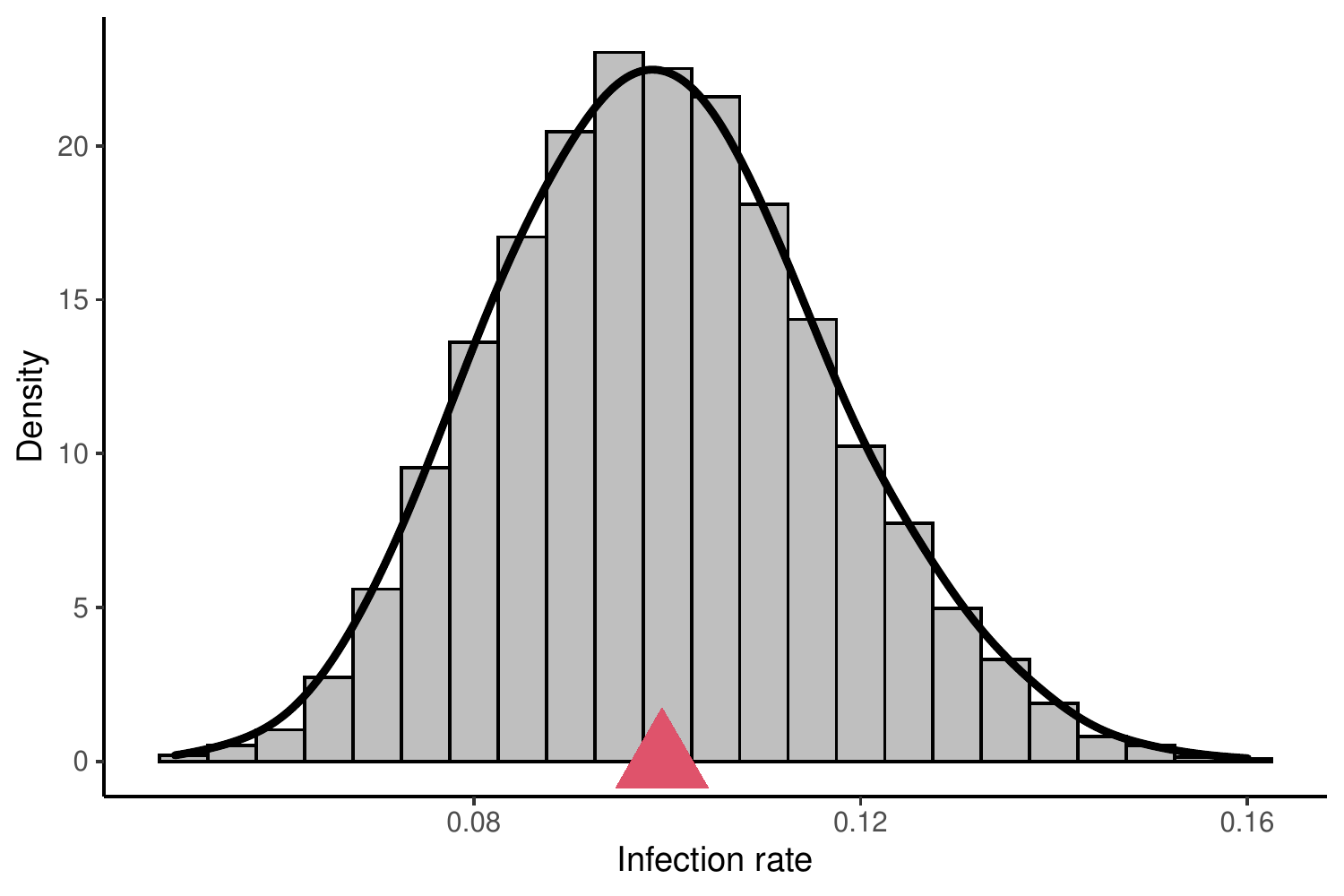}
    \includegraphics[scale=0.32]{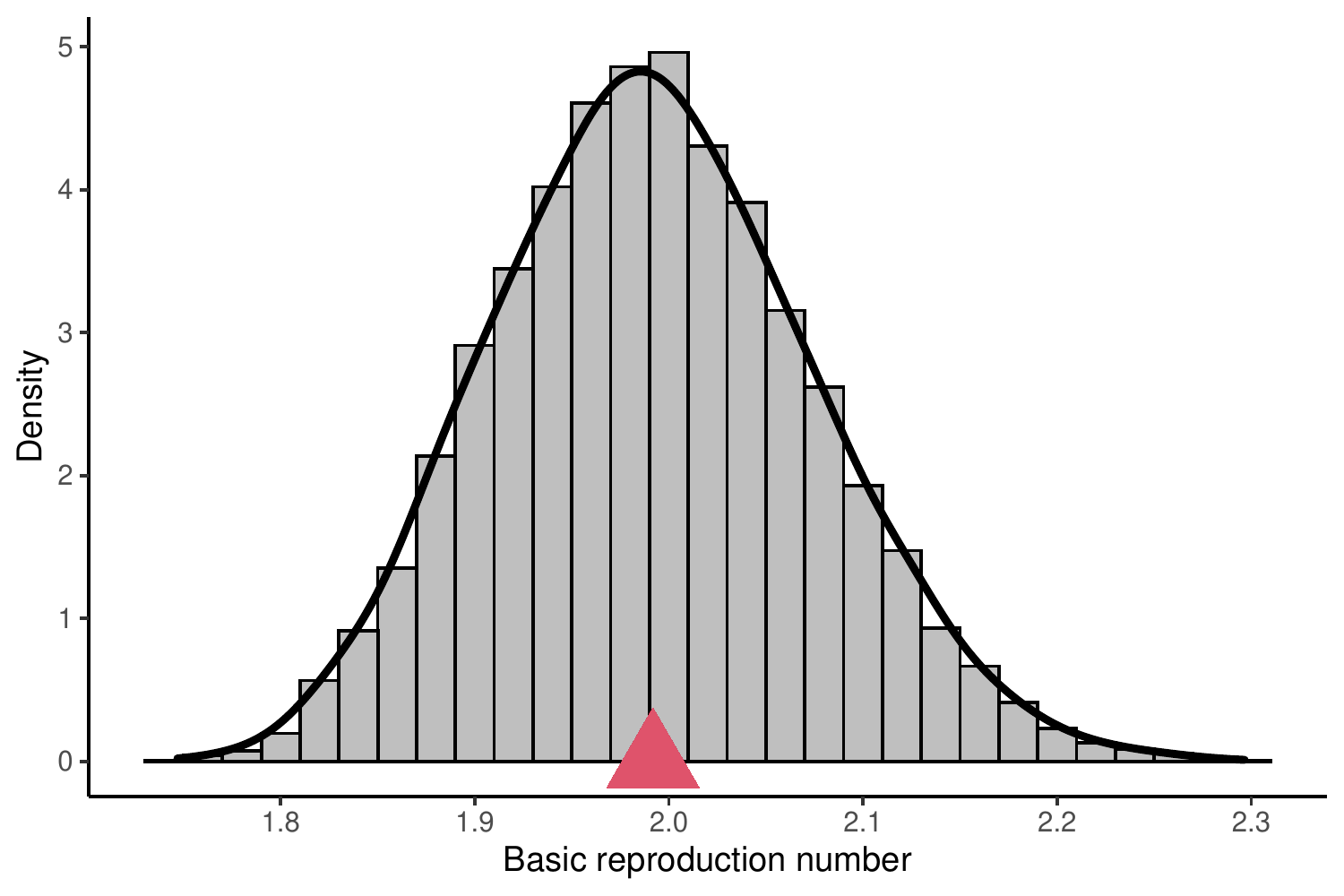}
    \includegraphics[scale=0.32]{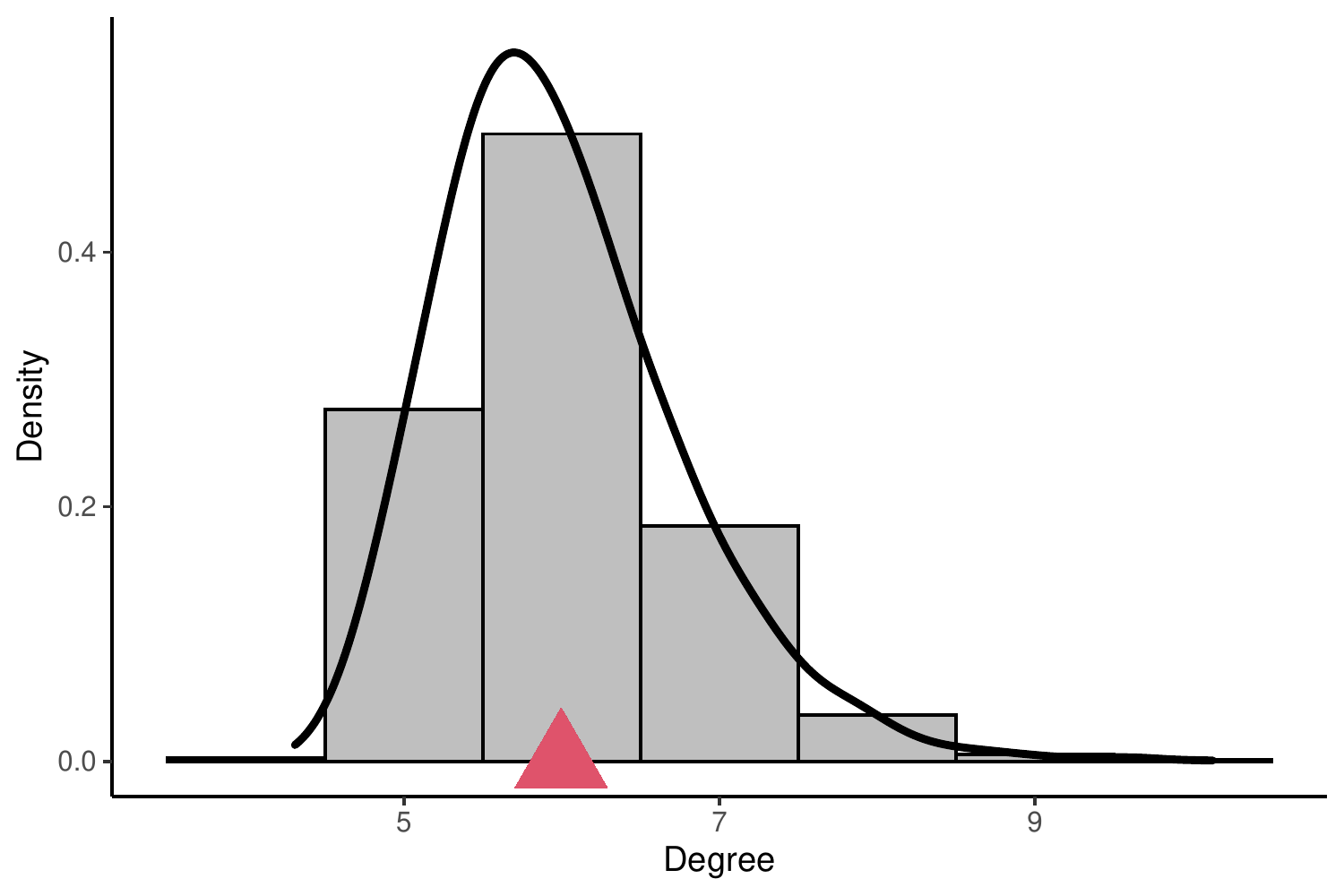}
    \includegraphics[scale=0.32]{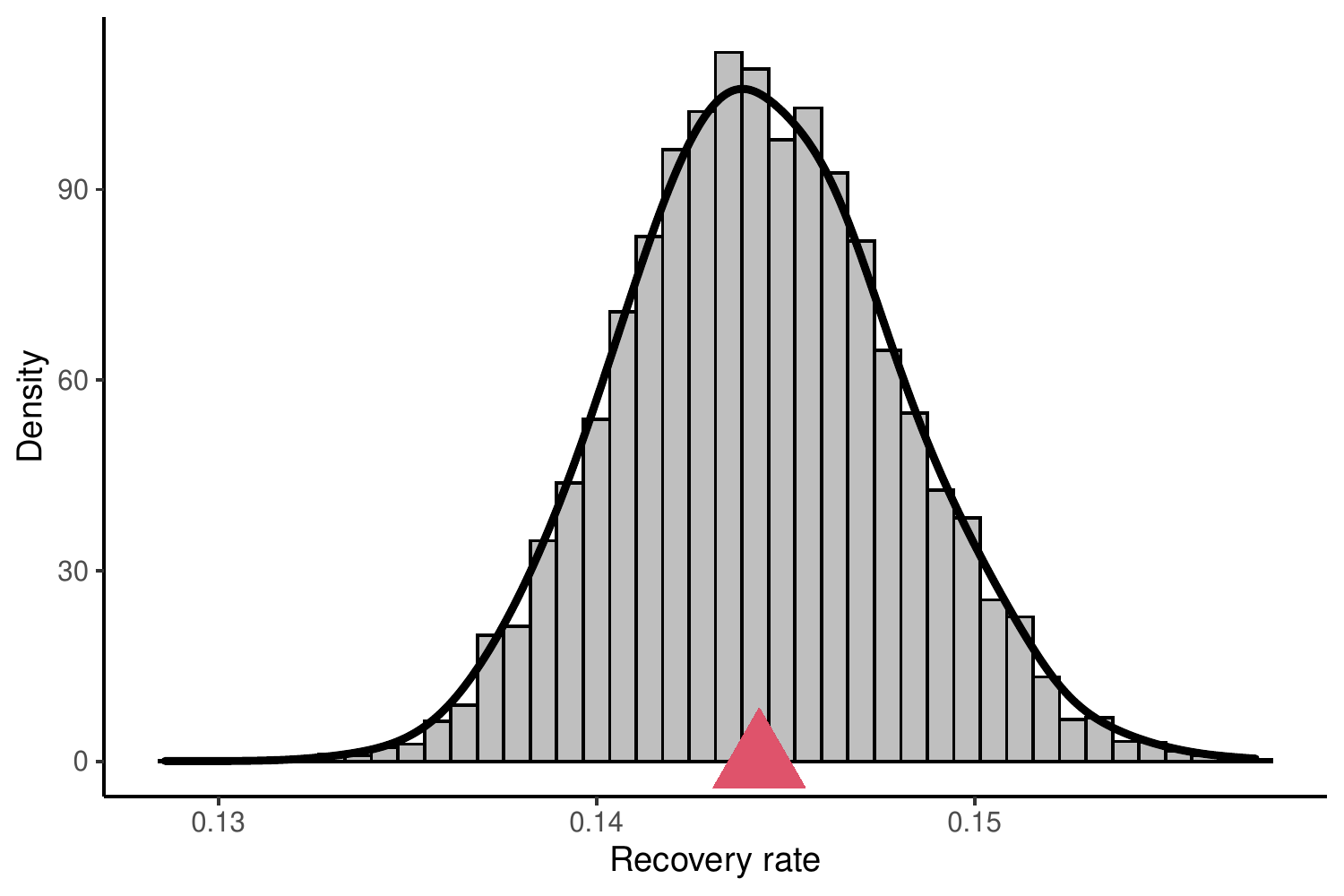}
    \includegraphics[scale=0.32]{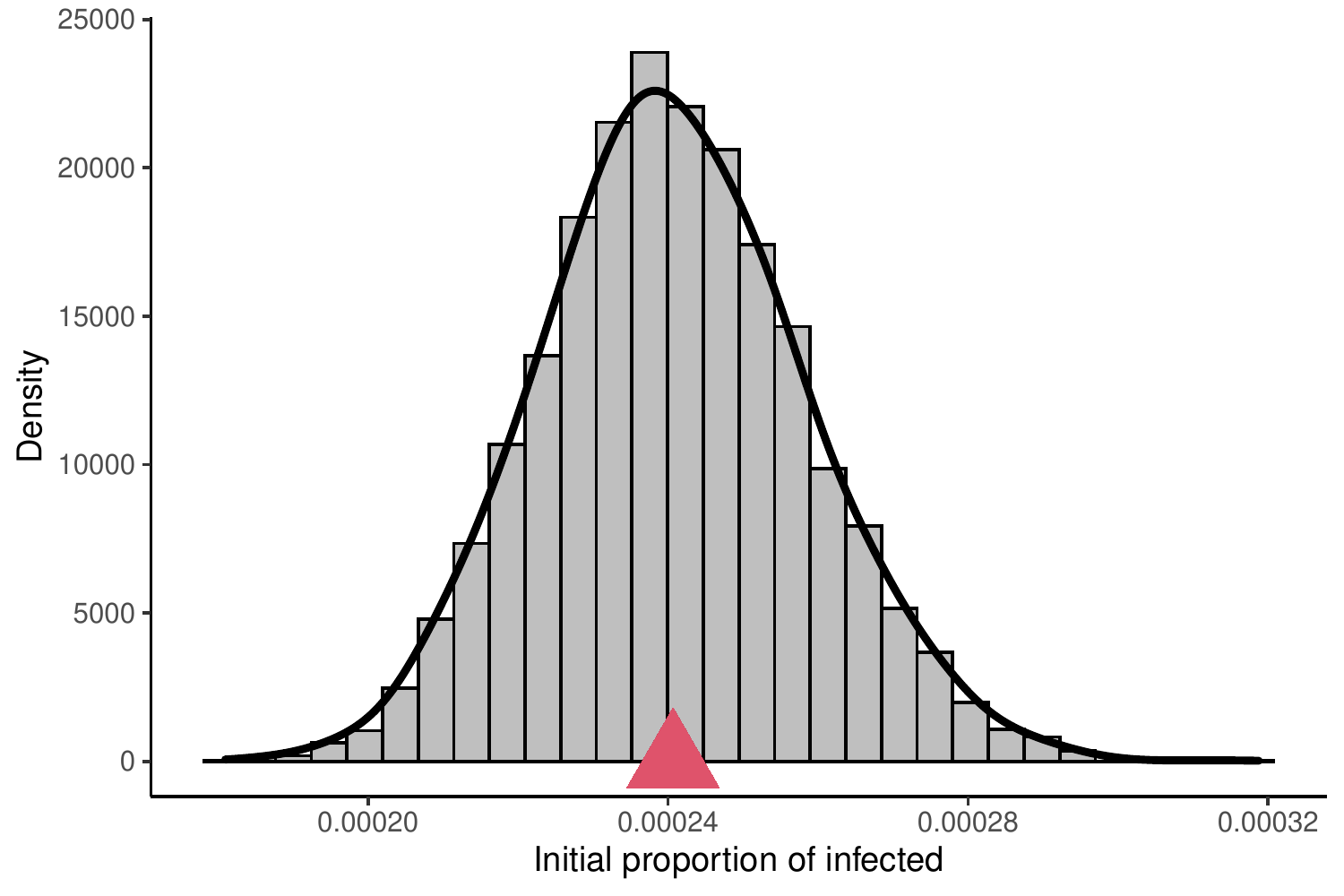}
    \includegraphics[scale=0.32]{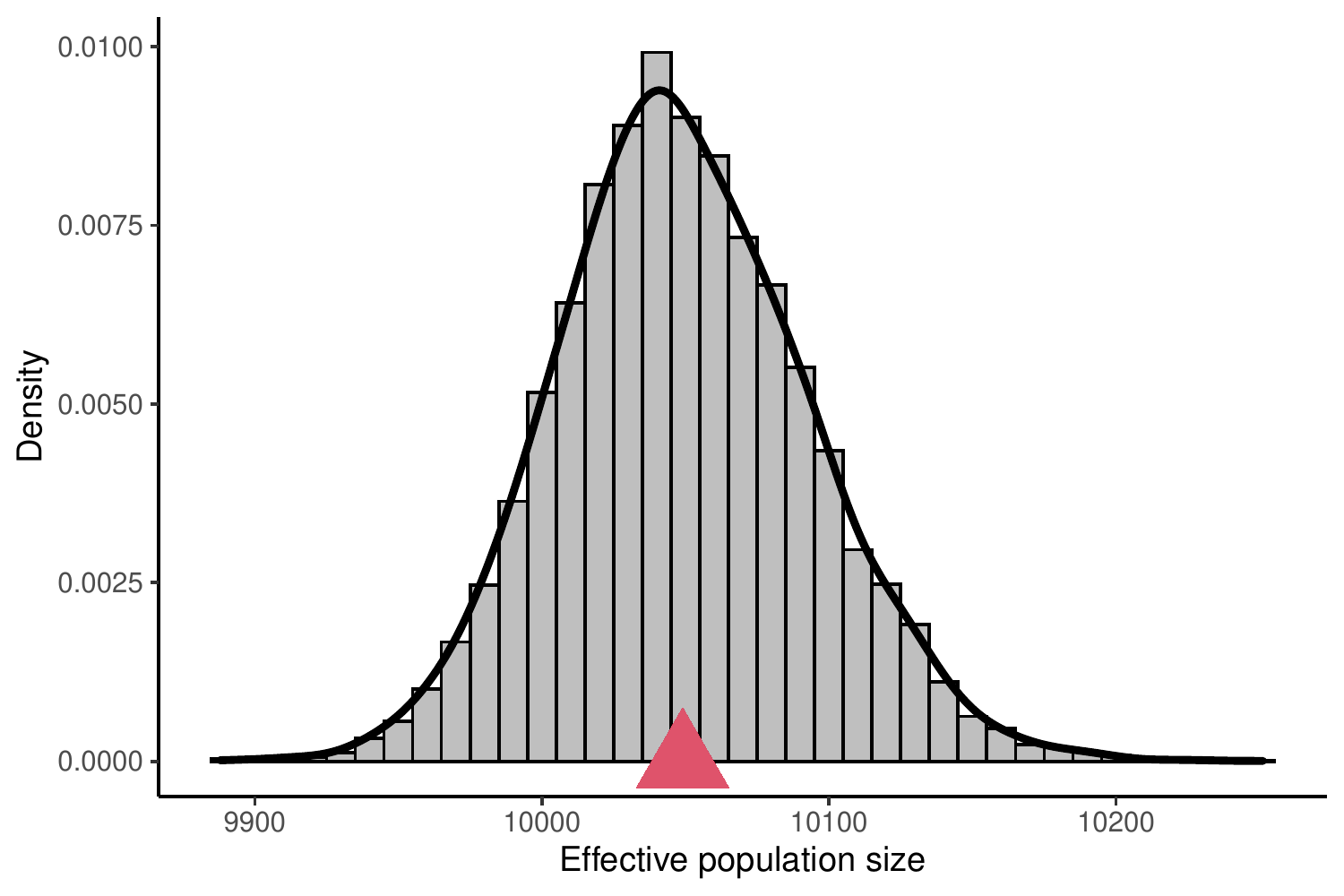}
    \caption{Posterior distributions of $(\tau, R_0, n, \gamma, \rho, n_T)$ using the DSA method on the synthetic data. The red triangles indicate the true values of the parameter. The means and the medians of the posterior distributions are $(0.0994, 1.992, 5.997, 0.144, 0.0002, 10049)$  and $(0.0989, 1.989, 5.891, 0.144, 0.0002, 10047)$, respectively. }
    \label{fig:dsa_exact_posteriors}
\end{figure}

Before describing the results of DSA on the synthetic data, we highlight that, unlike the MLE-based approach which either assumes or infers both population size and initial number of infected individuals (see also Section~\ref{subsec:MLE_system_size}), DSA inherently assumes an infinite population size (for both susceptible and infected individuals). Therefore, we do not infer the initial number of infected individuals. However, the ratio of initially infected to susceptible individuals, the parameter $\rho$, can be meaningfully inferred. In fact, having observed a finite number of infections in a given observation window $[0, T]$, DSA is also able to infer an \emph{effective population size} using the discount estimator \cite{KhudaBukhsh2020DSA,DiLauro2022NonMarkovDSA}:
\begin{align}
    n_T = \frac{k_T}{1-[S](T)}, 
\end{align}
where $k_T$ is the number of infections observed by time $T>0$. It should be noted that estimates of the effective population size depend on the observation time $T$, and could be substantially different from the true population size when applying the method to a real epidemic. Nevertheless, as evidenced by the posterior distributions of the parameters $(\tau, R_0, n, \gamma, \rho, n_T)$ shown in Figure~\ref{fig:dsa_exact_posteriors}, for this synthetic dataset, the method is able to infer the parameters well. The posterior distributions are unimodal, centred around the true values of the parameters. Here, at first random samples of individual infection  and recovery times (of size 5000 each) were constructed from the count dataset (one single trajectory of the Gillespie simulation) by drawing samples from appropriate uniform distributions (see Section~\ref{sec:dsa}). These random samples were then fed into the HMC scheme using four parallel Markov chains. Uninformative, flat priors were used. 

For this dataset, the parameter values estimated by both approaches are comparable. However, it is important to note that the two methods adopt two quite different likelihood constructions. Whilst the MLE-based approach relies on counts and the size of the population to construct the likelihood function, the DSA likelihood function only requires a random sample of infection times (and recovery times, if available). 
In other words, whilst the MLE-based approach assigns a likelihood to the epidemic trajectories, DSA identifies the probability laws of individual transfer times (infection and recovery times). These are often, even if censored, or truncated, more reliable and easily observed or derived statistical data than counts. For instance, even when we have partially observed count data on daily new infections, one can create a random sample of infection times (possibly censored/truncated).   Even when the entire population is \emph{not} monitored and only a set of randomly chosen individuals are followed through time and their transfer times are noted, the DSA methodology is still applicable. This advantage of DSA is particularly important when we fit the PW model to real epidemic data, which we do in the next section. 

\subsection{Inference from real-world data}

\subsubsection{System size and the MLE approach}\label{subsec:MLE_system_size}
\begin{figure}[h!]
	\center
	\includegraphics[scale=0.4]{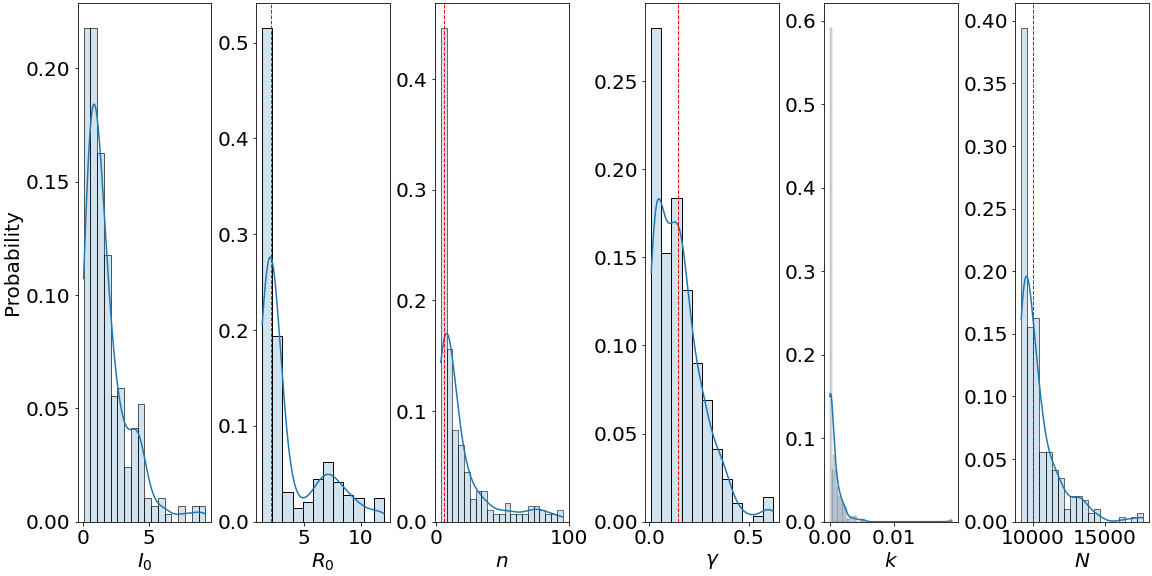}
	\caption{Inferring distributions for $[I_0, R_0, n, \gamma, k, N]$ for the stochastic realisations. The ground truth parameter values ($R_0$, $n$, $\gamma$ and $N$) are denoted by vertical dashed lines. Data shown correspond to 289 out of the 370 stochastic realisations (see detail in text).}
	\label{fig:Gillespie_MLE_N}
\end{figure}

In deploying the MLE approach to the above data, we used our knowledge of the true value of $N$. With real-world datasets, however, such information is typically not available. Whilst this is not an issue for DSA since it can infer an effective system size, it is for the MLE-based approach particularly in light of the unidentifiability issue discussed in \ref{sec::Identif}. In what follows, we infer the value of $N$ along with the other parameters, accepting that the increase in dimensionality of the parameter space will likely exacerbates unidentifiability. Here, we investigate the robustness of the inference process when inferring known parameters on the stochastic realisations presented in Section~\ref{sec::Gillespie}. The data presented in Figure~\ref{fig:Gillespie_MLE_N} result from the $289$ out of a possible $370$ realisations who satisfied the following conditions: (a) good fit (as quantified by the ratio $1.2$ to the smallest likelihood value $217.25$ obtained over the 370 realisations) -- this excluded 66 estimates, (b) reasonable $n$ (i.e., $n<500$ arbitrarily -- this excluded 13 estimates) and (c) reasonable $\gamma$ (i.e., $\gamma<1$ -- this excluded a further 2 estimates -- interestingly those estimates had very large $N$, specifically 26038.74 and 30168.98 but still showed very low nLL (244.02 and 228.3 respectively). The median values for the 6 parameters were: $I_0=1.256792$, $R_0=2.11$, $n=8.84$, $\gamma=0.129$, $k=0.00002$ and $N=9877.83$. These values are reasonably close to the theoretical values ($I_0=1$, $R_0=2$, $n=6$, $\gamma=0.14$ and $N=10000$) which is encouraging. In particular, the percentage error in $N$ is under $1.5\%'$ (For reference, the percentage error for DSA on the same data is in the order of $0.01\%$). Nevertheless, as shown by Figure~\ref{fig:Gillespie_MLE_N}, there is substantial variance in the estimates including significantly higher values of both $N$ and $R_0$ (e.g., 70 estimates have $R_0>4$) despite excellent fits. 

\begin{figure}[h!]
	\center
	\includegraphics[scale=0.4]{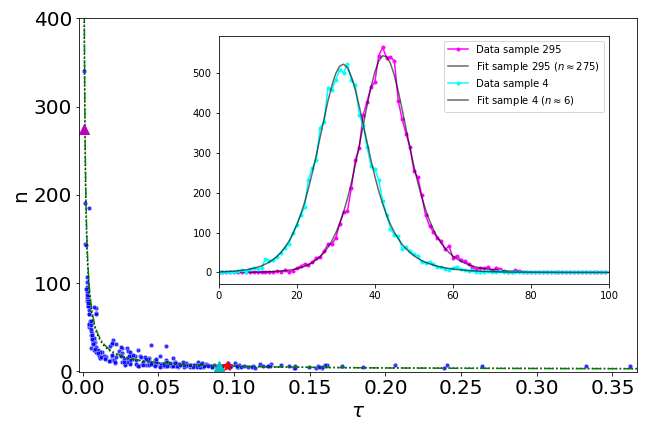}
	\caption{Main panel: Scatter plot of the parameter estimates on the $n$, $\tau$ plane with the two unidentifiability curves calculated as per Eqs.~\ref{eq:tau19} (dotted line), and ~\ref{eq:tau20} (dashed line). The star denotes the true values, i.e., true $n$ and calculated value of $\tau$ given true values of $R_0$ and $n$. Only those estimates who did not provide a good fit, as per the criterion above) were excluded, resulting in $304$ surviving estimates. Inset: Empirical data and fit for two stochastic realisations corresponding to the triangles in the main panel with two significantly different inferred degree $n$ (see detail in text).}
	\label{fig:Gillespie_UC_N}
\end{figure}

To illustrate this point, we plotted the estimates on the $(\tau, n)$ plane (see Figure~\ref{fig:Gillespie_UC_N}) and confirmed that they conform to the unidentifiability curve previously identified. The inset shows two stochastic realisations and the corresponding fits with one fit producing an estimate for the degree $n$ close to the true value (6) and one producing an estimate magnitudes of order larger (275). As shown by the Figure (as well as the likelihood values), the fits are equally excellent. Inferred parameters for the data with the expected degree were: $I_0=2.38$, $R_0=2.18$, $n=6.05$, $\gamma=0.111$ and $N=9689.76$, i.e., close to the ground truth data. In contrast, the inferred parameters for the data with the large degree were: $I_0=0.24$, $R_0=10.56$, $n=274.92$, $\gamma=0.024$ and $N=9281.53$. 

\subsubsection{FMD data}
Unlike DSA, the MLE approach only provides a single point estimate. This makes it difficult to provide a meaningful comparison of the two methods. To mitigate this issue, we repeated the MLE inference process $100$ times, each time using a different set of initial conditions. To construct the equivalent of a posterior, we included all parameter values obtained in each of the $100$ times, provided the nLL was sufficiently close to the best nLL over the $100$ rounds. The number of estimates excluded for each dataset will be reported but highlights the fact that the search algorithm can get stuck in very sub-optimal local minima. 

\begin{figure}[h!]
	\center
	\includegraphics[scale=0.4]{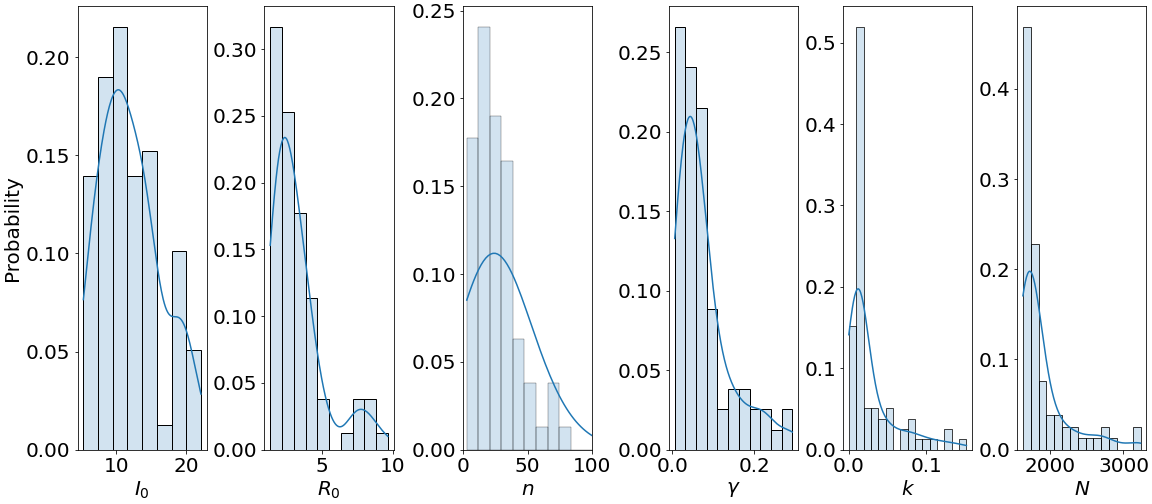}
	\caption{Distributions of $[I_0, R_0, n, \gamma, k, N]$ using MLE on the FMD data using $100$ rounds of inference with different initial conditions. The median values are listed in Table~\ref{tab:empirical_MLE_stats} and compared with the DSA approach in Section~\ref{subsec:MLEvsDSAcomparison}. Five estimates for which $n>100$ ($154$, $156$, $279$, $294$ and $368$) were excluded from the figure (but not the statistics) for improved readability of the histogram.}
	\label{fig:FMD_MLE_N}
\end{figure}

Histograms of inferred parameters for the FMD dataset using the MLE approach are shown in Figure~\ref{fig:FMD_MLE_N}. $11$ out of $100$ estimates were excluded because of an anomalous outcome of the inference process. The estimates with the lowest nLL are $I_0=10.54$, $R_0=2.58$, $n=153.67$, $\gamma=0.0723$, $k=0.010$, and $N=1817.2$. There is quite a bit of dispersion around the parameters, with fairly fat tails. For example, whilst the  median for $R_0$ (2.71, see Table~\ref{tab:empirical_MLE_stats}) is relatively close to the best estimate, we also observe some fairly large values (in fact $10$ out of $100$ estimates were excluded because of $R_0>10$). The best and median estimate for $N$ was $1817$ and $1747$ respectively. This number is very likely implausible as well more than $2000$ farms will have been involved in the epidemic, but see DSA results below. Likewise the inferred average degree seems far overinflated. The value of $\gamma \eqsim 0.07$ implies 14 days for the infection period. Note that previous studies, see \cite{DiLauro2022NonMarkovDSA} for example, have reported a mean of $10.2$ days. Importantly, the fits are good with all (accepted) estimates showing a very narrow range of nLL values (from 233.03 to 248.67 with a mean of 236.31 and a std of 4.06). This once again provides evidence of the fact that the MLE approach ascribes a likelihood to the trajectory produced by the inferred parameters rather than to the parameters themselves.  

\begin{figure}[h!]
    \centering
    \includegraphics[scale=0.33]{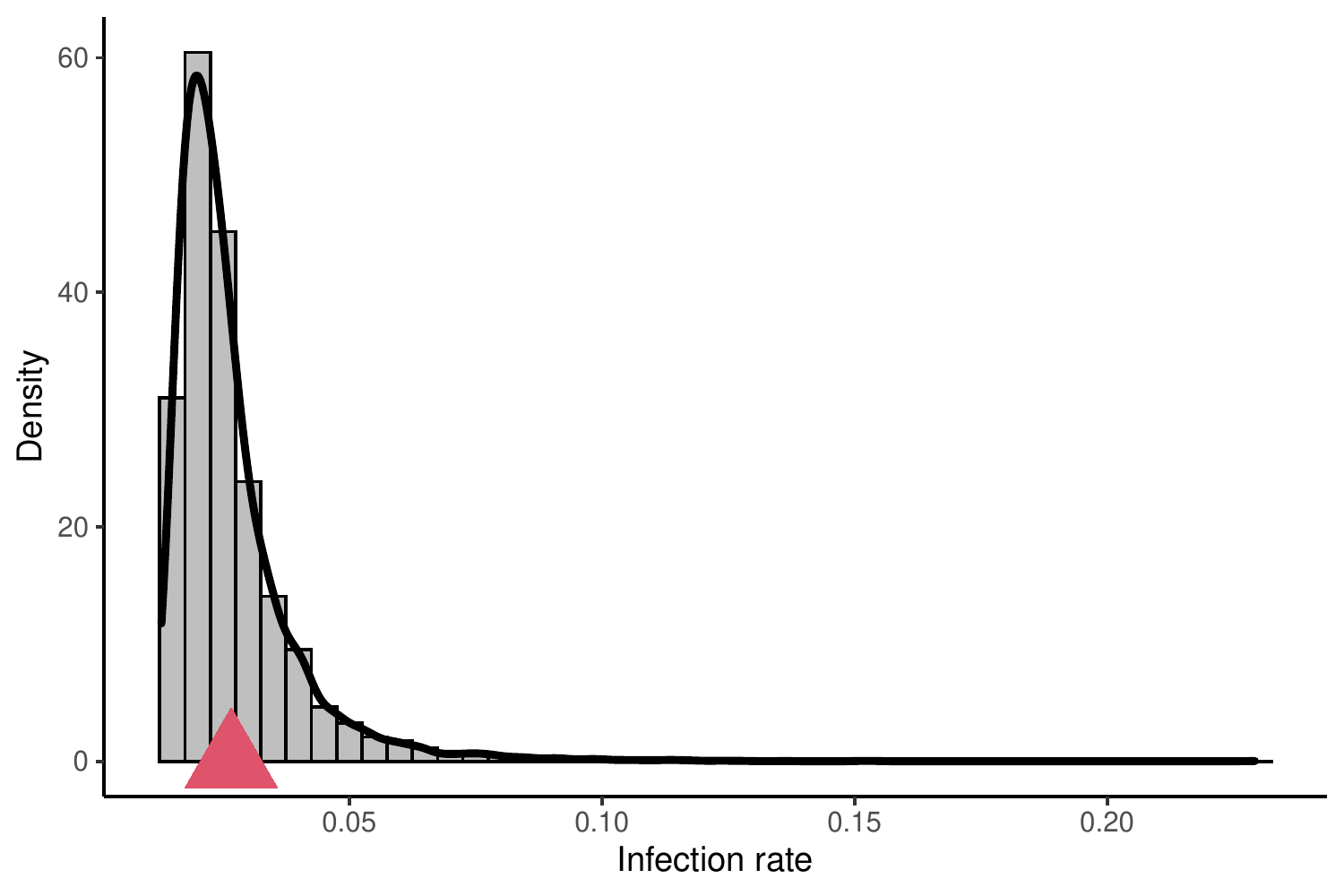}
    \includegraphics[scale=0.33]{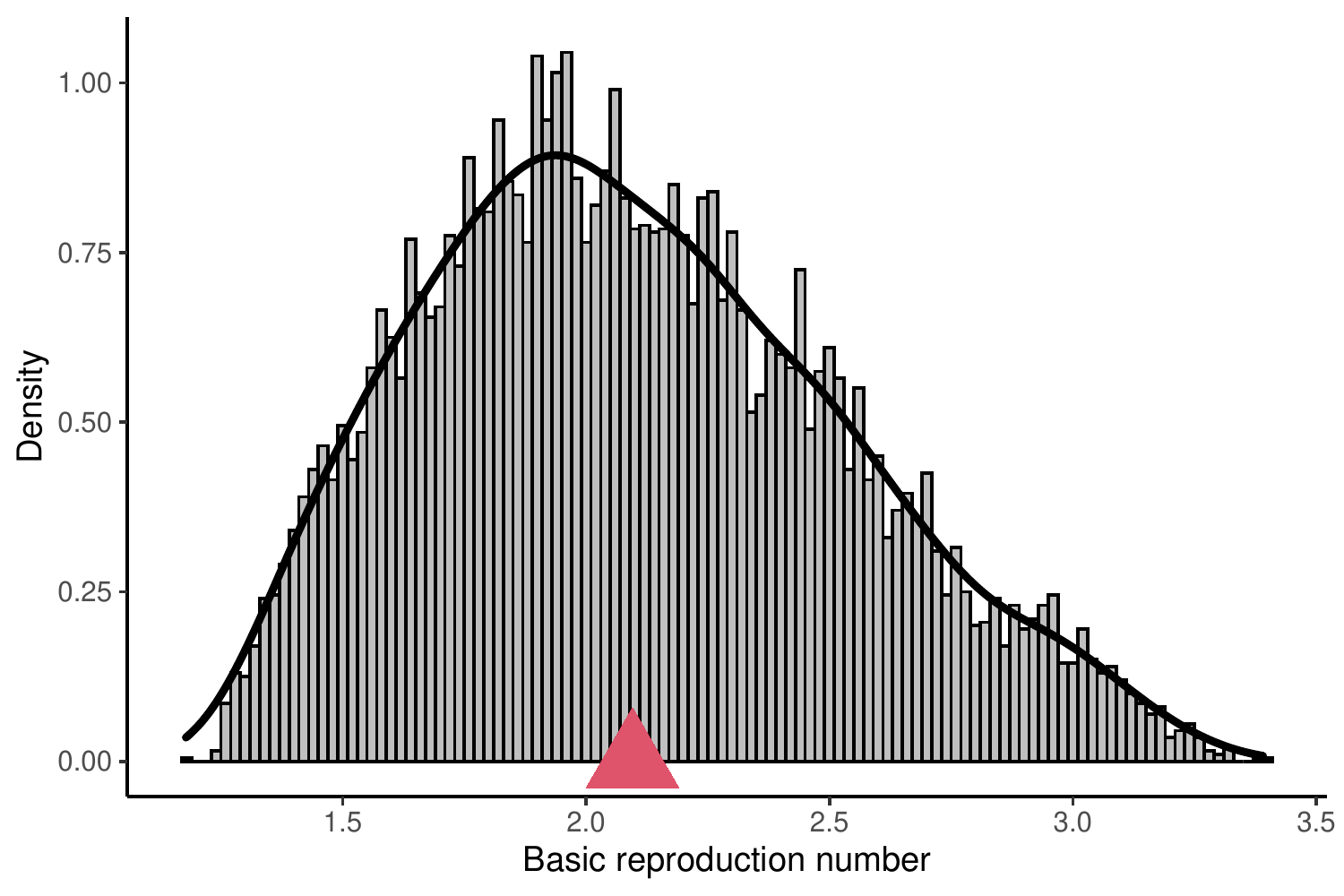}
    \includegraphics[scale=0.33]{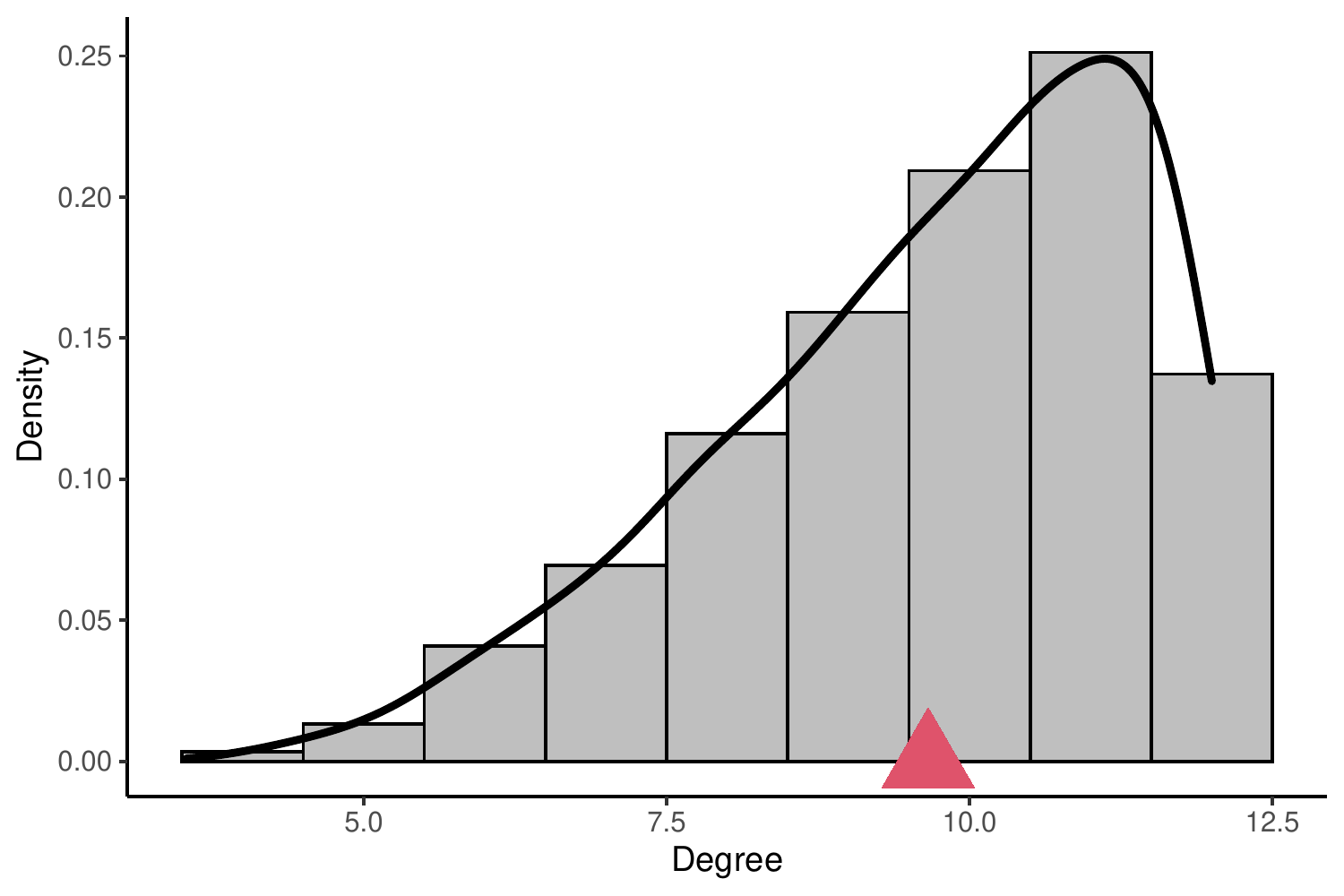}
    \includegraphics[scale=0.33]{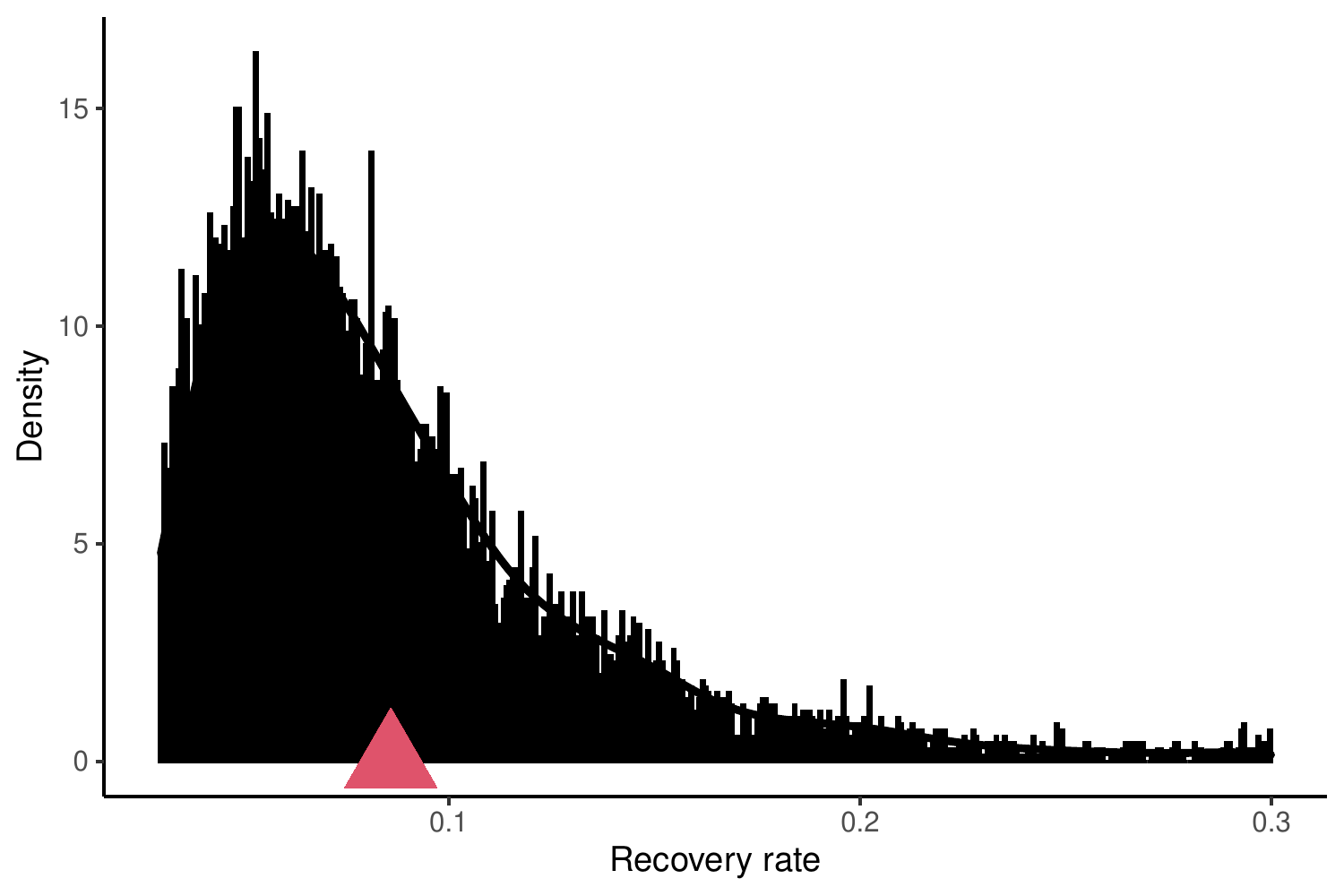}
    \includegraphics[scale=0.33]{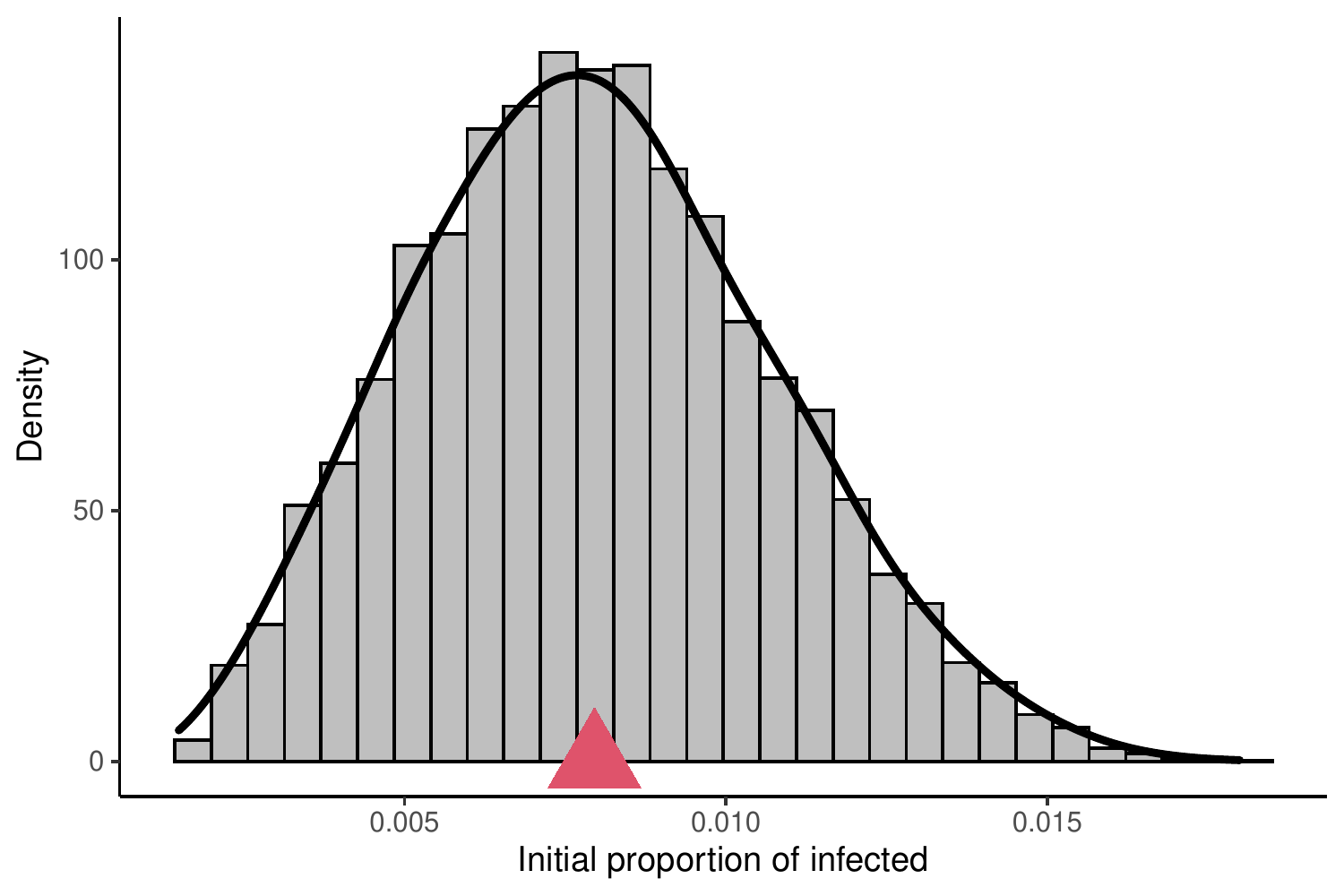}
    \includegraphics[scale=0.33]{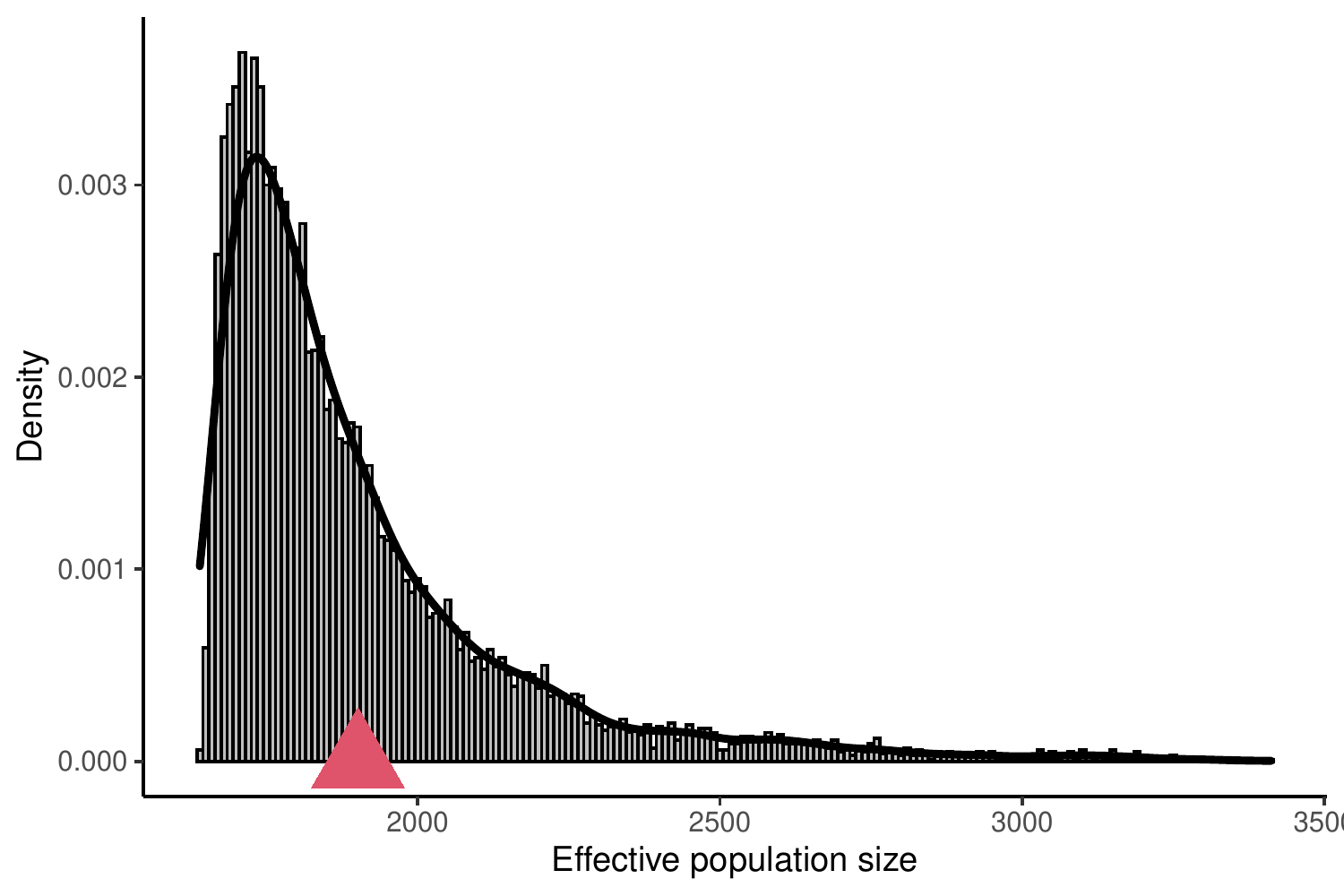}
    \caption{Posterior distributions of $(\tau, R_0, n, \gamma, \rho, n_T) $ using DSA on the FMD dataset. The red triangles indicate the means of the posterior distributions. The means and medians of the posterior distributions are  $(0.0266, 2.095, 9.659, 0.0859, 0.0079, 1901)$ and $(0.0233, 2.054, 9.982, 0.0737, 0.0078, 1819)$, respectively.}
    \label{fig:fmd_dsa_posteriors}
\end{figure}

The posterior distributions obtained by DSA method on the FMD dataset are shown in Figure~\ref{fig:fmd_dsa_posteriors}. It is important to note that, unlike with the MLE approach, these results were obtained when using an informative prior, an exponential distribution with mean $10.2$ days, for the $\gamma$ parameter following on the analysis in \cite{DiLauro2022NonMarkovDSA}. The posterior distributions are unimodal. The mean estimates are consistent with previously reported values, for example in \cite{DiLauro2022NonMarkovDSA}. Interestingly, and as with the MLE approach, the estimated effective population size is less than 2000. This is not to be confused with the number of farms, however (see brief explanation in Section~\ref{subsec:MLEvsDSAcomparison}).  

\subsubsection{H1N1-N18234}
\begin{figure}[h!]
	\center
	\includegraphics[scale=0.4]{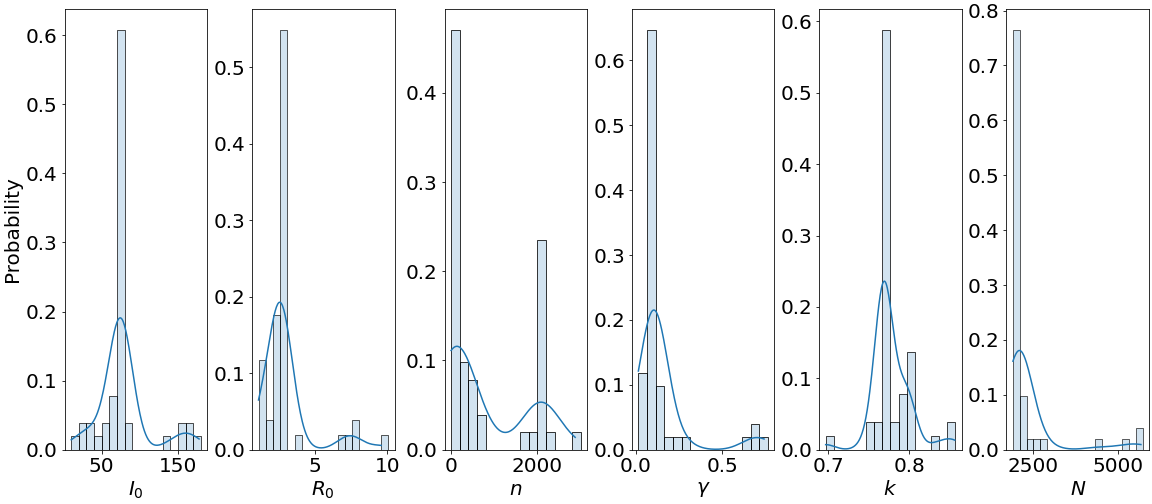}
	\includegraphics[scale=0.4]{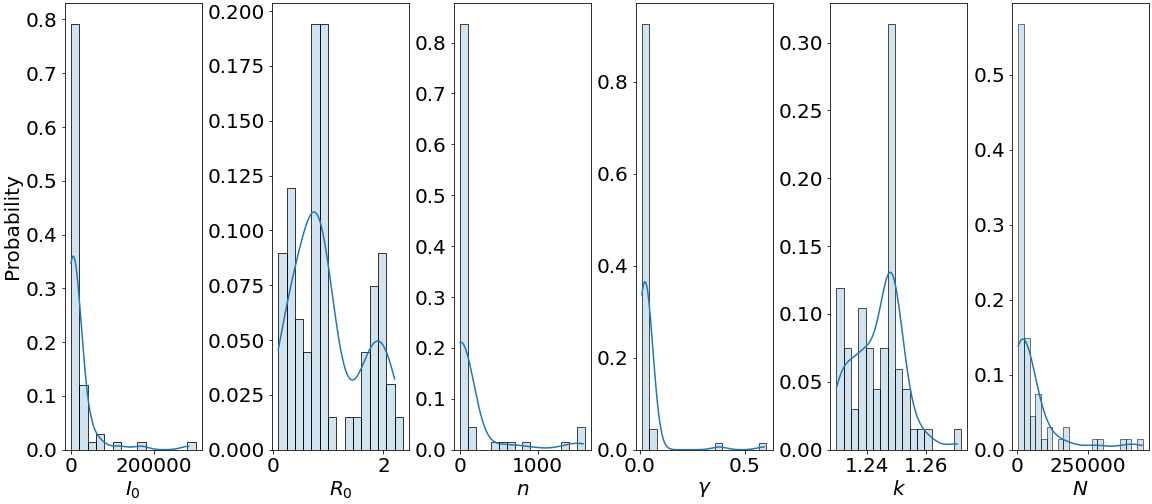}
	\caption{Distributions of $[I_0, R_0, n, \gamma, k, N]$ using MLE on the H1N1 data (with horizon restricted to 42, top panel and full data, bottom panel) using $100$ rounds of inference with different initial conditions. The median values are listed in Table~\ref{tab:empirical_MLE_stats}.}
	\label{fig:ah1n1_MLE_N}
\end{figure}

The A(H1N1) dataset presents an interesting challenge as it has a long persistent tail with visible stochastic effects. We therefore present two sets of results: one where we infer parameters on the full dataset (i.e., including the tail) and one when we restrict to $T=42$. Figure~\ref{fig:ah1n1_MLE_N} shows the results of the MLE-based approach for both scenarios. As clearly evidenced by the bottom right panel of Figure~\ref{fig:MLE_vs_DSA_fits}, when the full horizon is considered, the fits are poor, the noisy tail seemingly obfuscating the true trajectory of the epidemic. Not surprisingly, the parameter estimates appear meaningless and highly variables from one round of inference to the other despite similar nLL (see Table~\ref{tab:empirical_MLE_stats}). When restricting to $T=42$, the fits are good and the parameter estimates are slightly better behaved albeit with not unimodal and with implausibly large $n$ considering the inferred population size $N$. In fact, only $51$ out of $100$ parameter estimates survived once we excluded 3 estimates for being poor fits, 13 for excessive values of $R_0$ ($>10$) and 33 estimates for excessive value of $\gamma>1$. Interestingly, we note the high value of $k$ inferred in both scenarios, with MLE correctly recognising the high dispersion of the counts.   

\begin{figure}[h!]
    \centering
    \includegraphics[scale=0.33]{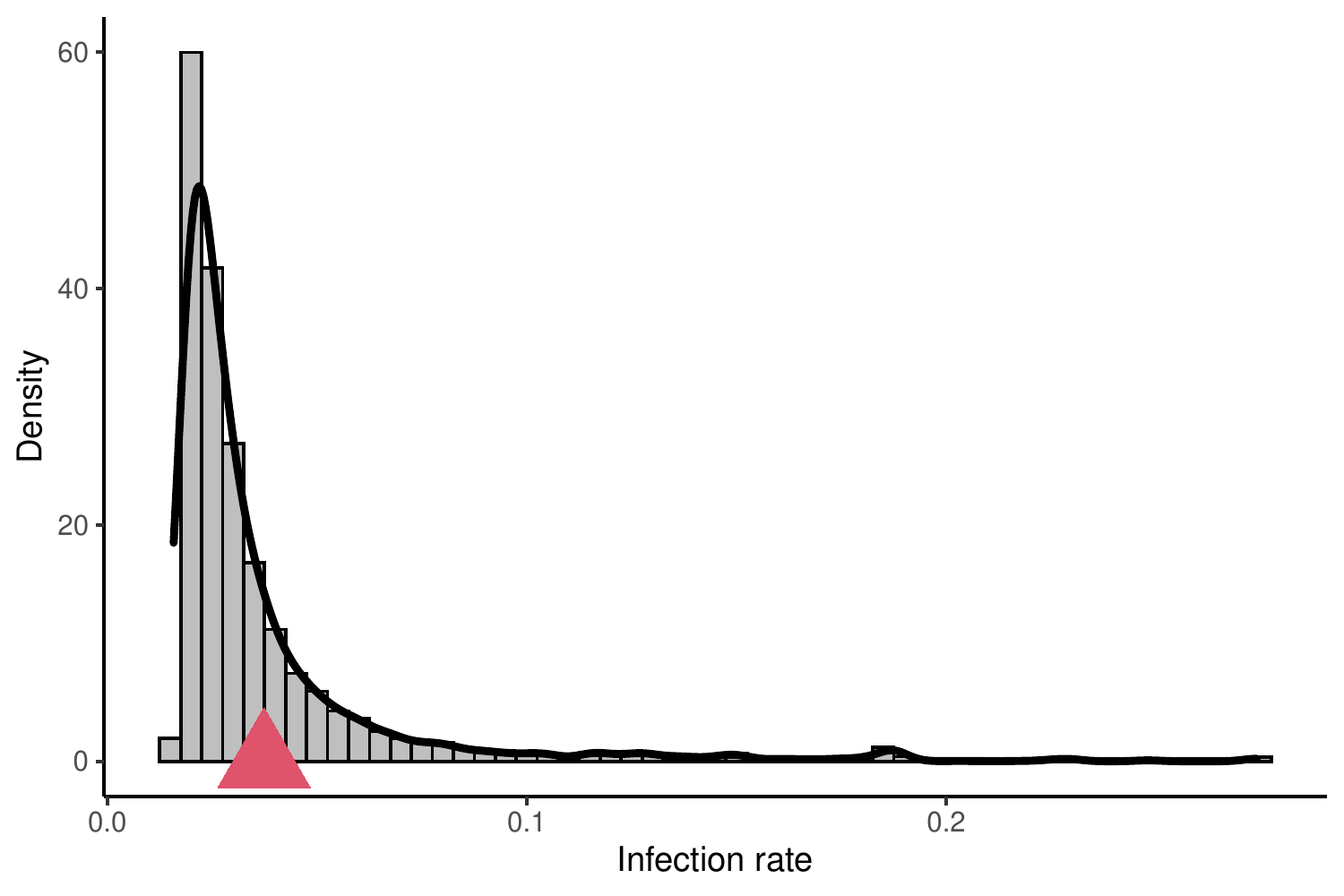}
    \includegraphics[scale=0.33]{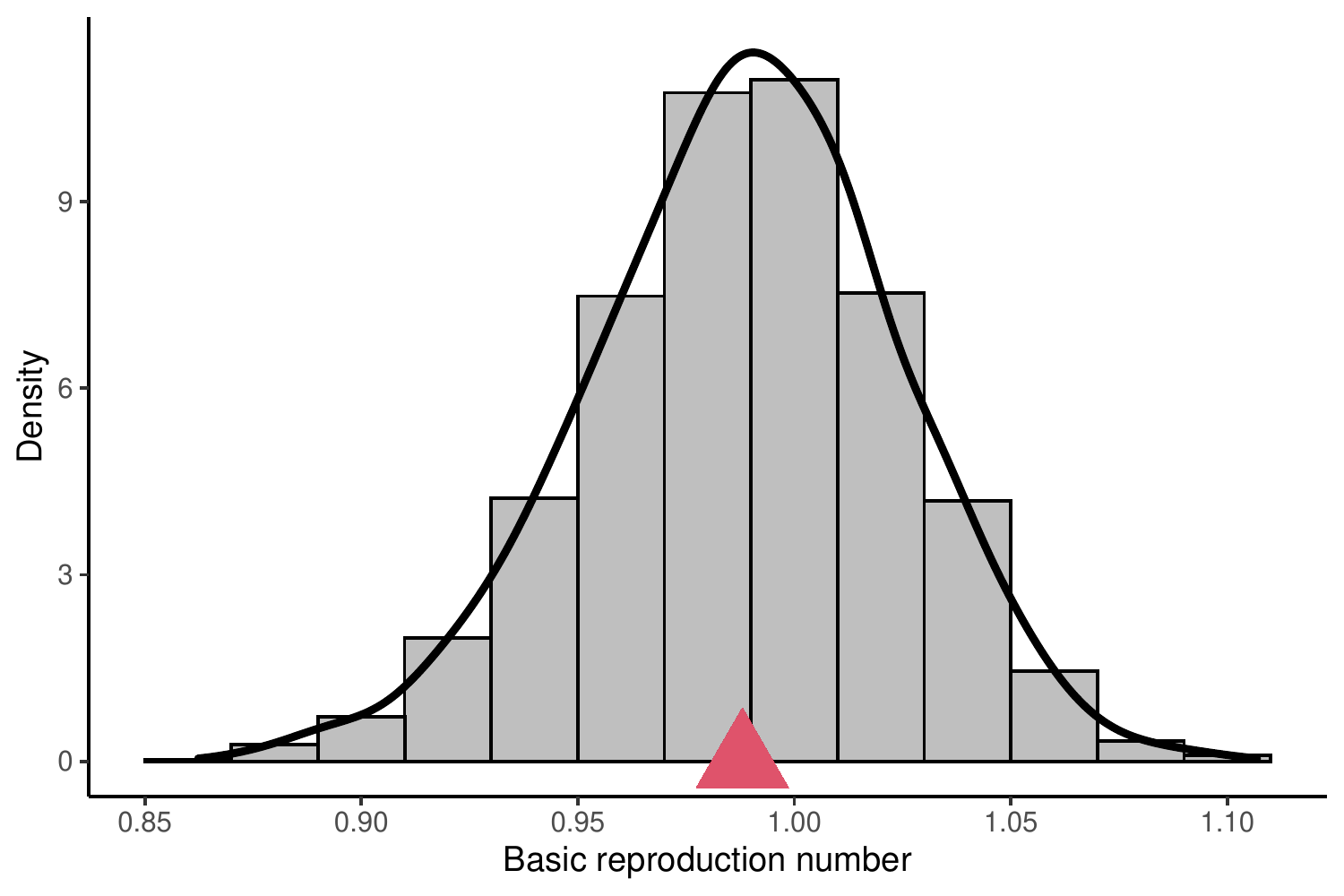}
    \includegraphics[scale=0.33]{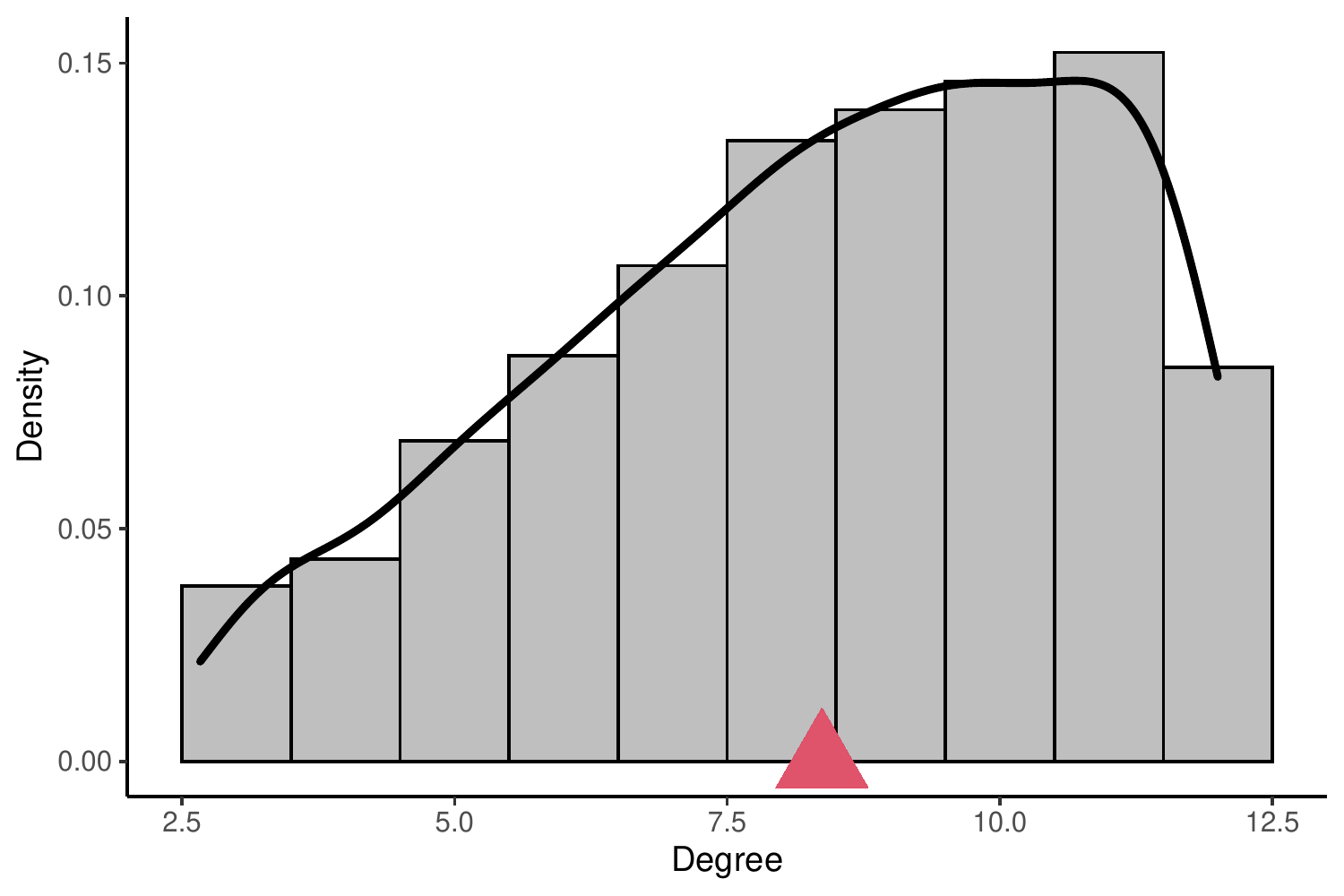}
    \includegraphics[scale=0.33]{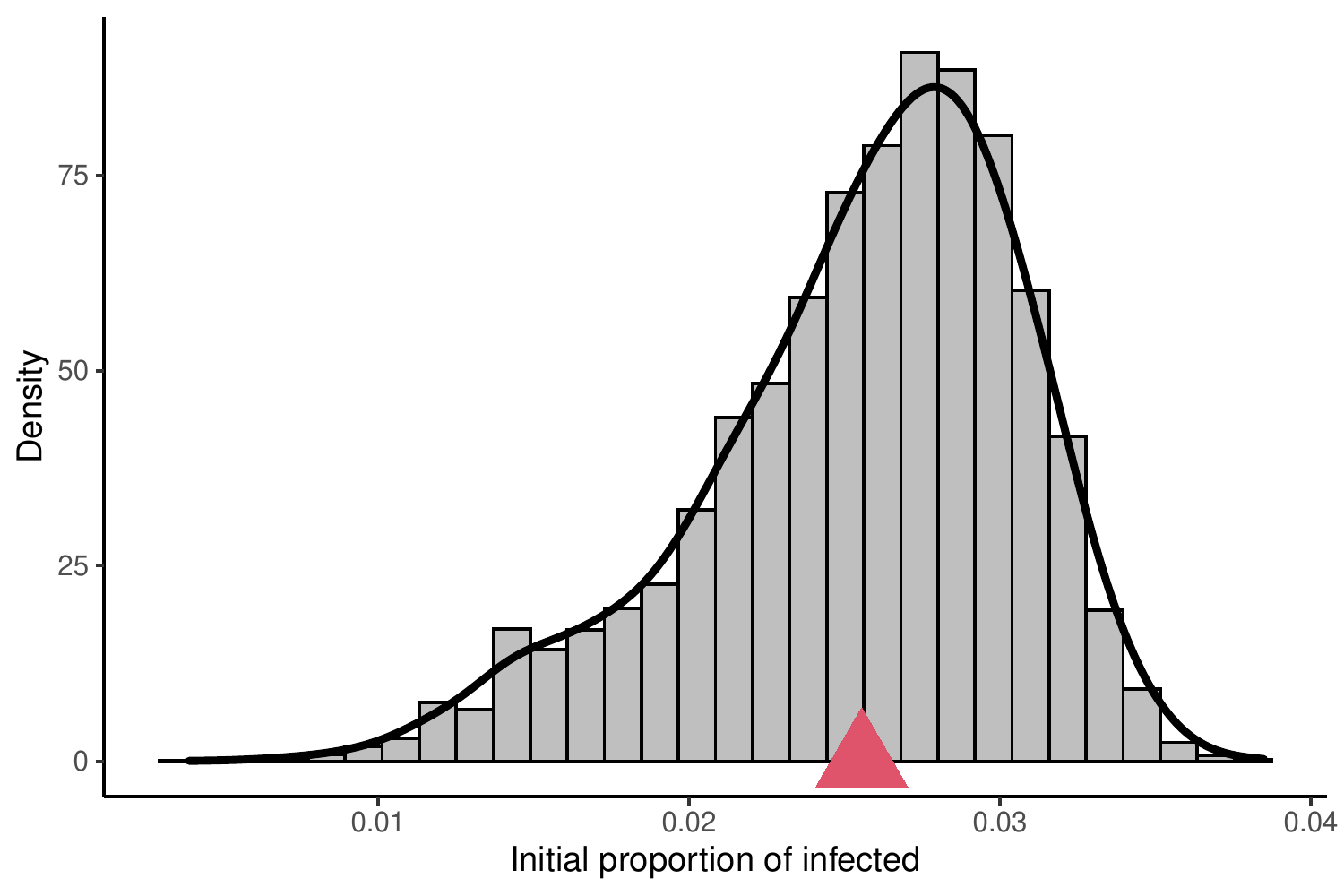}
    \includegraphics[scale=0.33]{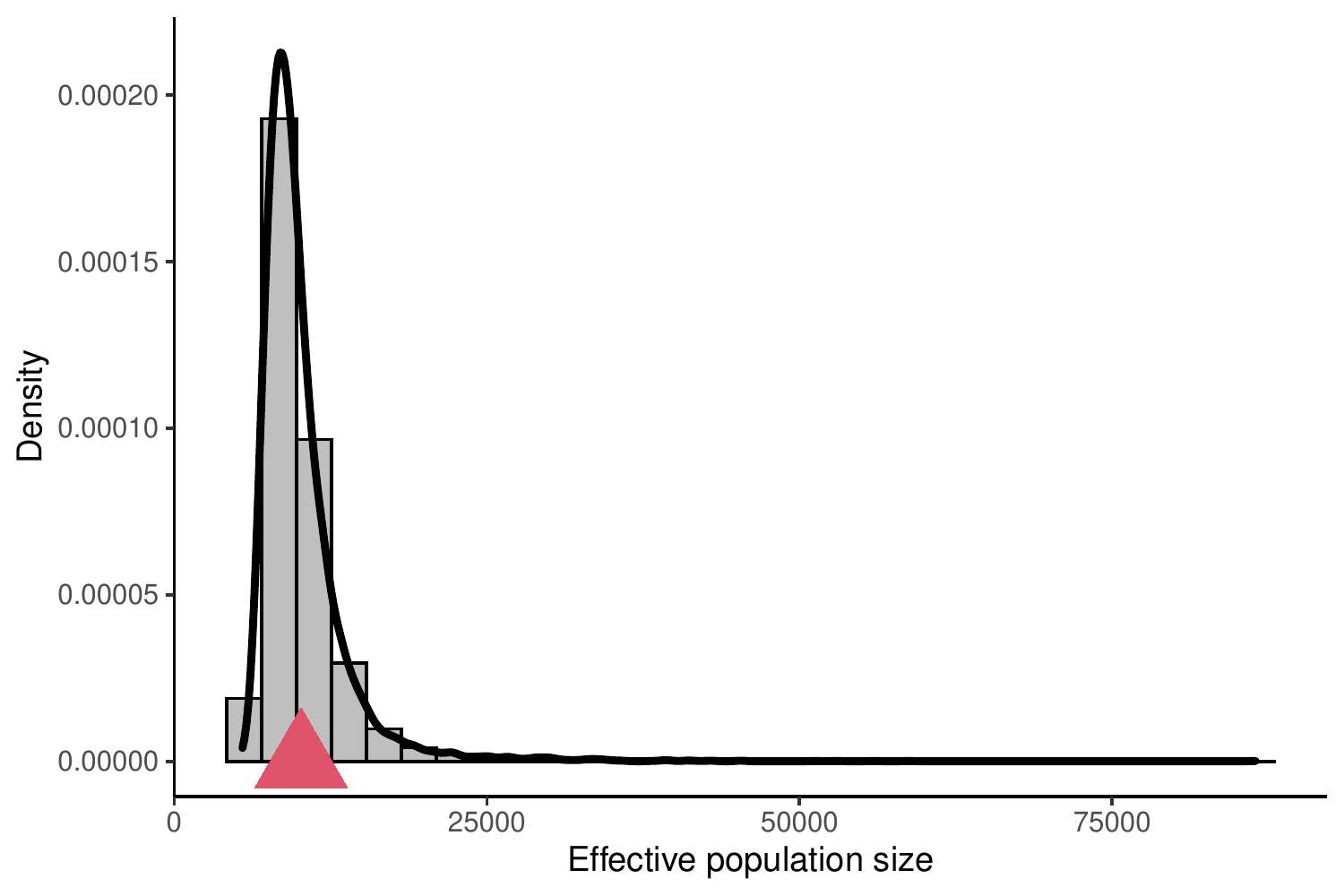}
    \caption{Posterior distributions of $(\tau, R_0, n, \rho, n_T)$ using DSA on the full A(H1N1) outbreak data. 
    The means and the medians of the posterior distributions are $(0.0373, 0.9880, 8.369, 0.0255, 10146)$ and $(0.0269, 0.9892, 8.665, 0.0264, 9286)$, respectively.}
    \label{fig:ah1h1_dsa_posterior}
\end{figure}

\begin{figure}[h!]
    \centering
    \includegraphics[scale=0.33]{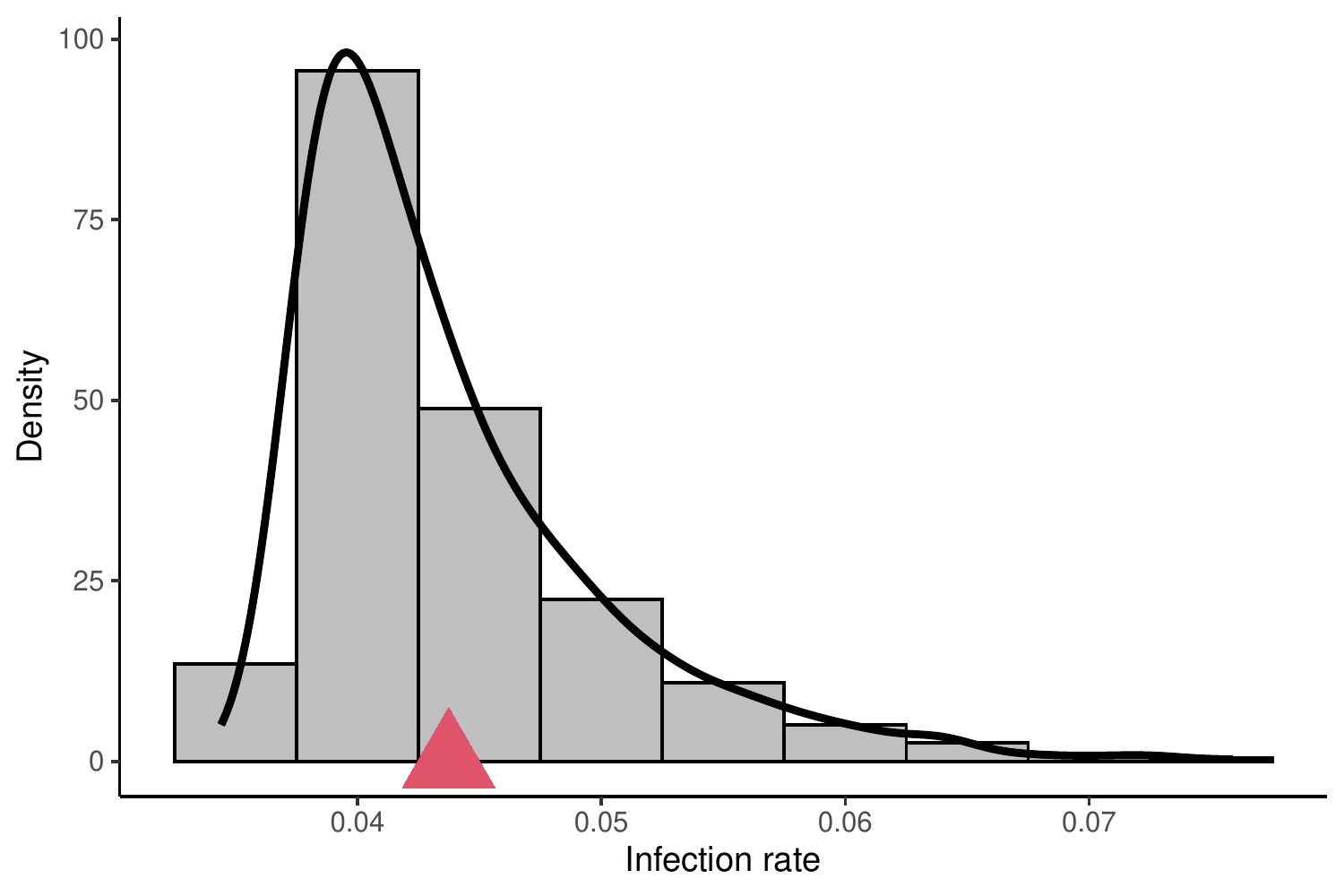}
    \includegraphics[scale=0.33]{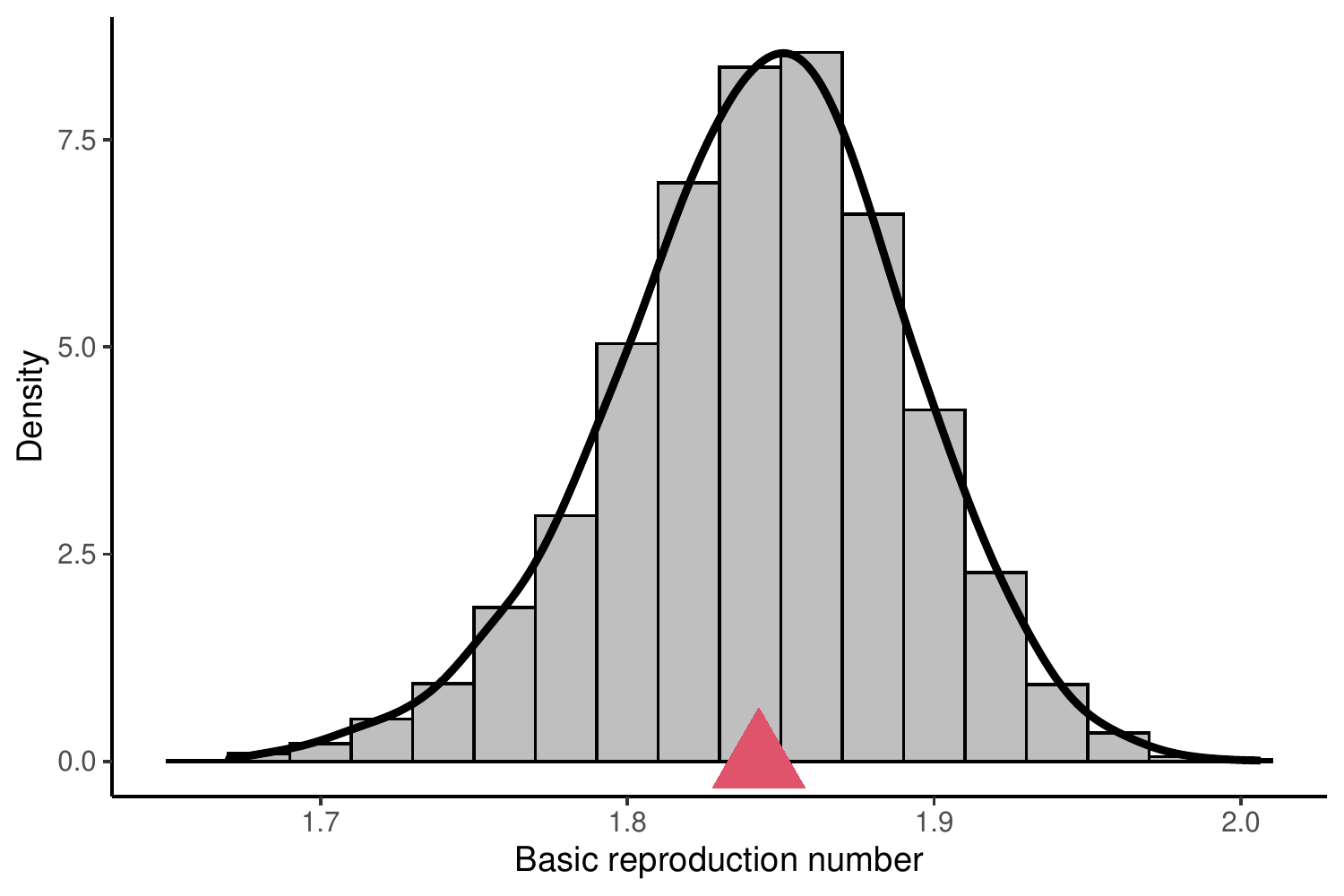}
    \includegraphics[scale=0.33]{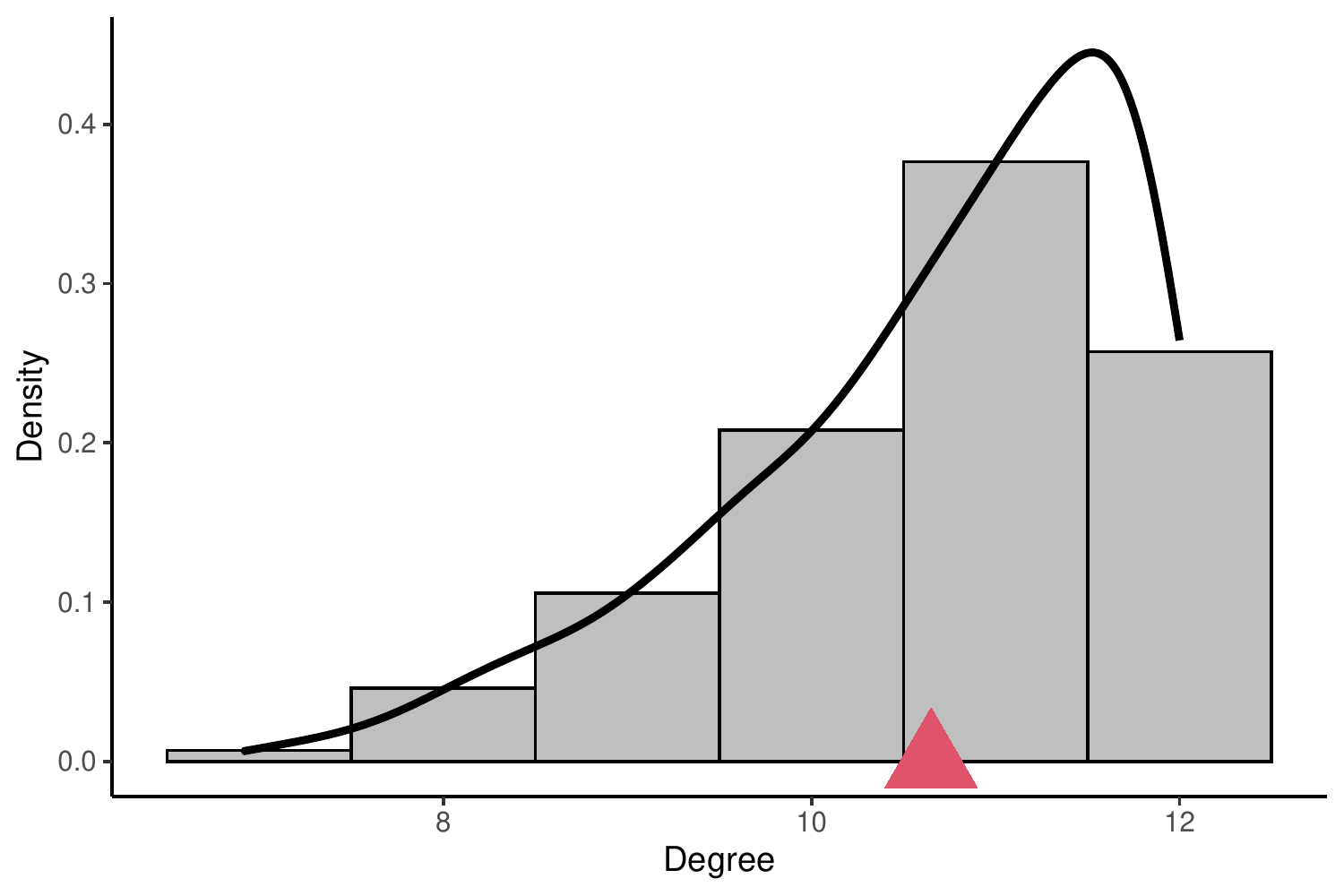}
    \includegraphics[scale=0.33]{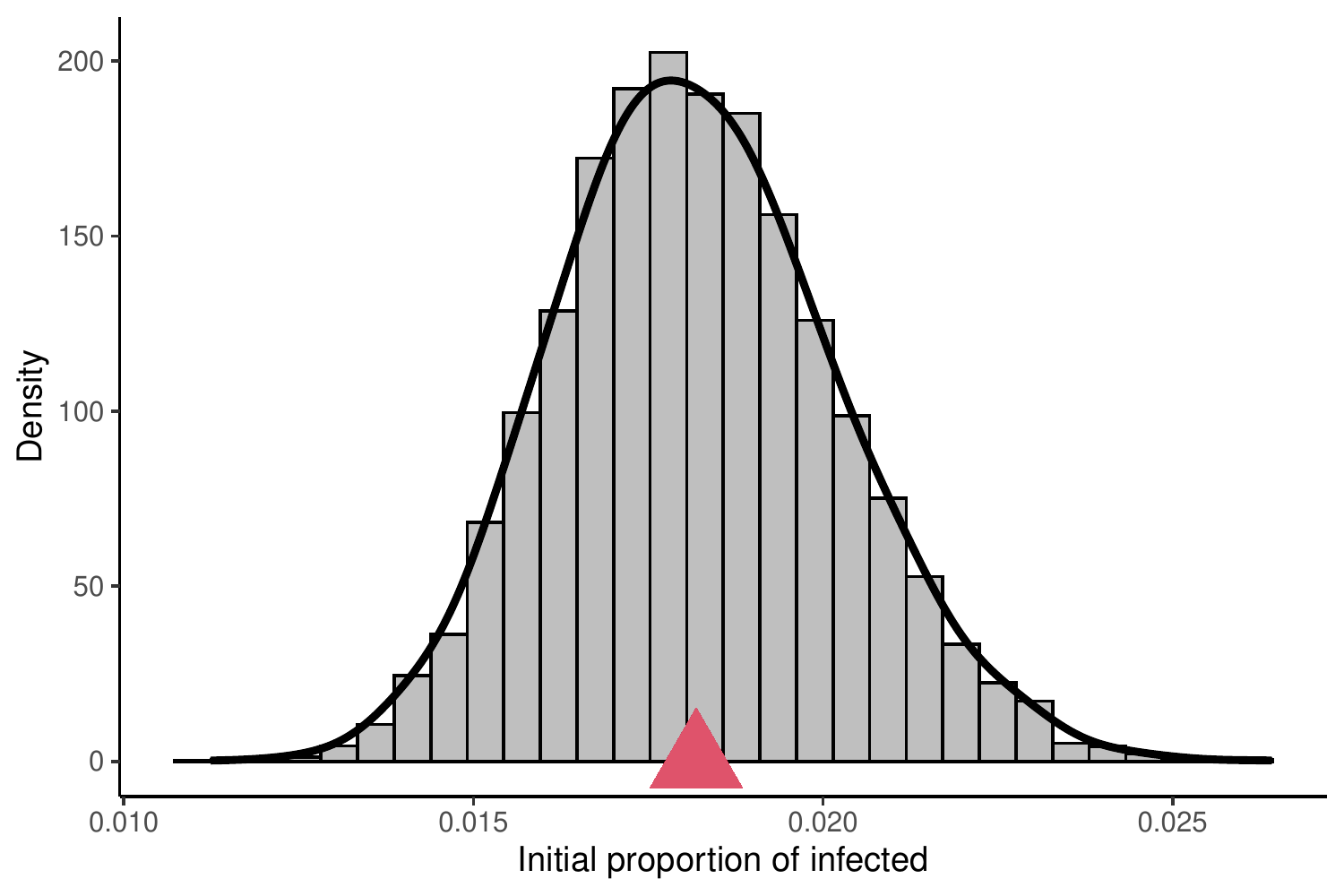}
    \includegraphics[scale=0.33]{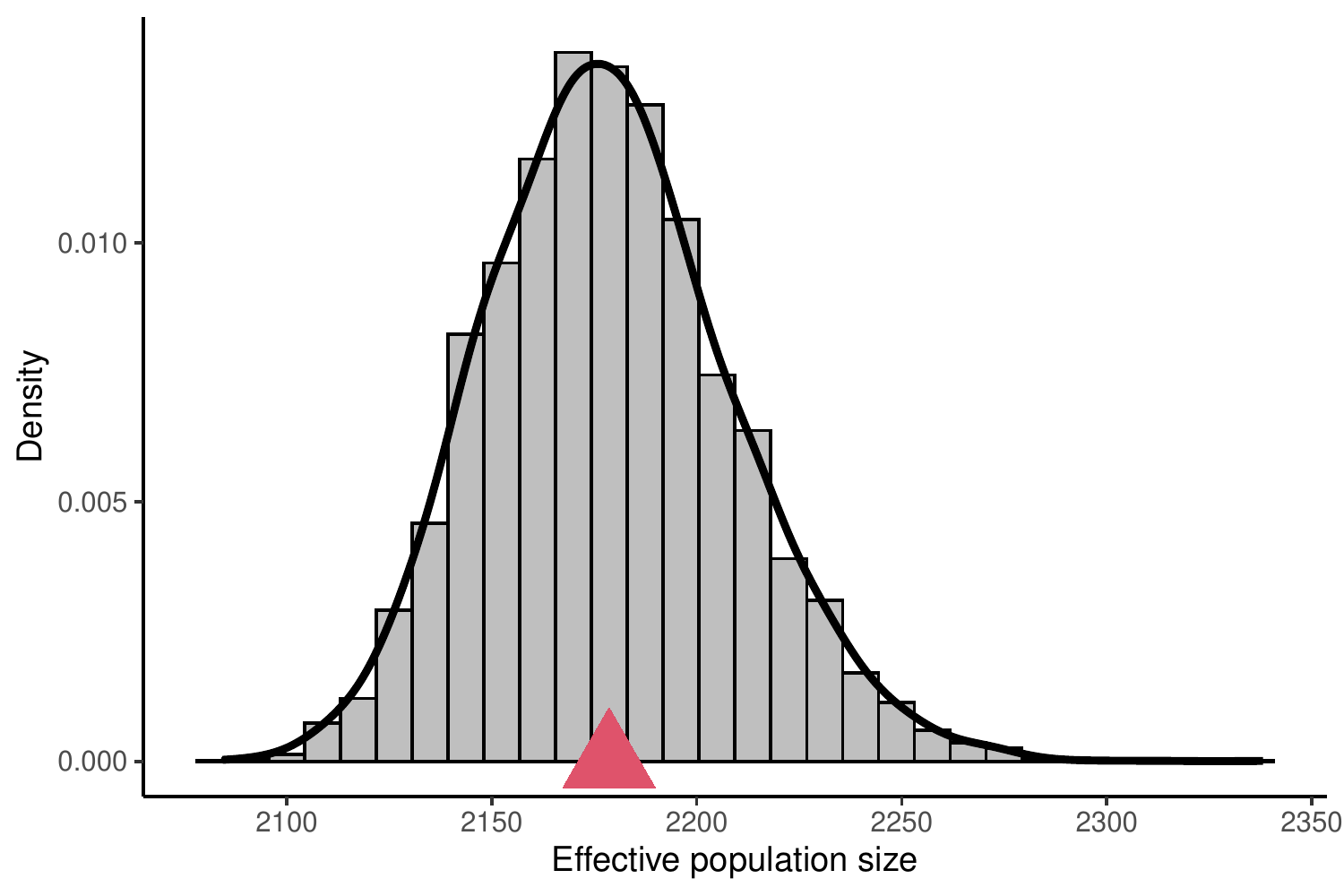}
    \caption{Posterior distributions of $(\tau, R_0, n, \rho, n_T)$ using DSA on the A(H1N1) outbreak data restricted to time horizon $T=42$. 
    The means and medians of the posterior distributions are $(0.0437, 1.843, 10.650, 0.0189, 2179)$ and $(0.0418, 1.845, 10.908, 0.0189, 2177)$, respectively.}
    \label{fig:small_ah1h1_dsa_posterior}
\end{figure}

When deploying DSA, once again, a prior was used for $\gamma$ ($\gamma^{-1}=5.5$) based on published literature (see \cite{schwartz2014estimating, KhudaBukhsh2020DSA}). Figures~\ref{fig:ah1h1_dsa_posterior} and ~\ref{fig:small_ah1h1_dsa_posterior} show the posterior distributions of the parameters $(\tau, R_0, n, \rho, n_T)$ based on the full and partial data respectively. As with the MLE-based approach, when fitting to the full data, the DSA fit is poor, and in fact, very similar to that of the MLE approach (see bottom right panel of Figure~\ref{fig:MLE_vs_DSA_fits}). When removing the noisy tail of the data, the quality of inference improves significantly with both MLE and DSA producing near identical fits (bottom left panel of Figure~\ref{fig:MLE_vs_DSA_fits}). However, unlike with the FMD dataset, the inferred parameters are quite different although interestingly the ML-estimated population size and the DSA effective size are very similar (see Table~\ref{tab:empirical_MLE_stats}). 

\subsubsection{COVID-19 in India}
\begin{figure}[h!]
	\center
	\includegraphics[scale=0.4]{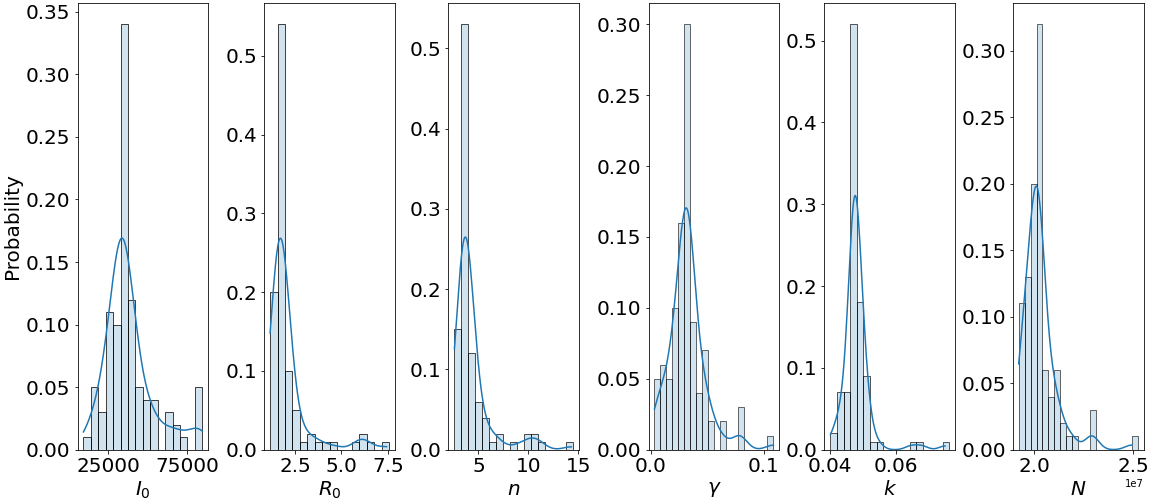}
	\caption{Distributions of $[I_0, R_0, n, \gamma, k, N]$ using MLE on the COVID-19 dataset using $100$ rounds of inference with different initial conditions. The median values are listed in Table~\ref{tab:empirical_MLE_stats}.}
	\label{fig:covid_MLE_N}
\end{figure}

Figure~\ref{fig:covid_MLE_N} shows the histograms of the estimates obtained by the ML-based approach on the final dataset. Here, unlike with the previous dataset, there was high consistency between estimates over the $100$ rounds with no exclusions needed. Curiously, this homogeneity of results is associated with an apparent mismatch between the fitted model and the data, as shown by the top right panel in Figure\ref{fig:MLE_vs_DSA_fits}. 

\begin{figure}[h!]
    \centering
    \includegraphics[scale=0.33]{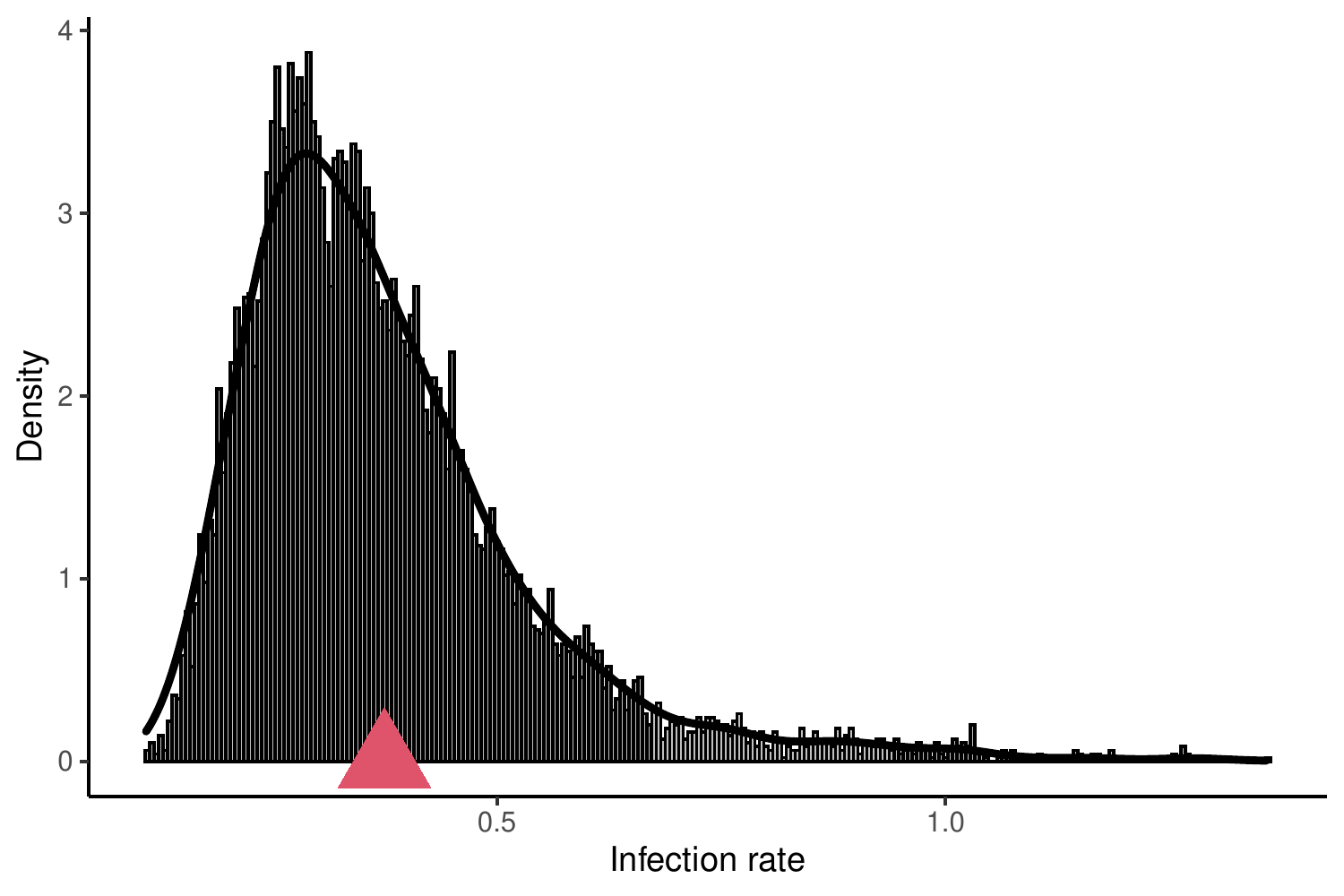}
    \includegraphics[scale=0.33]{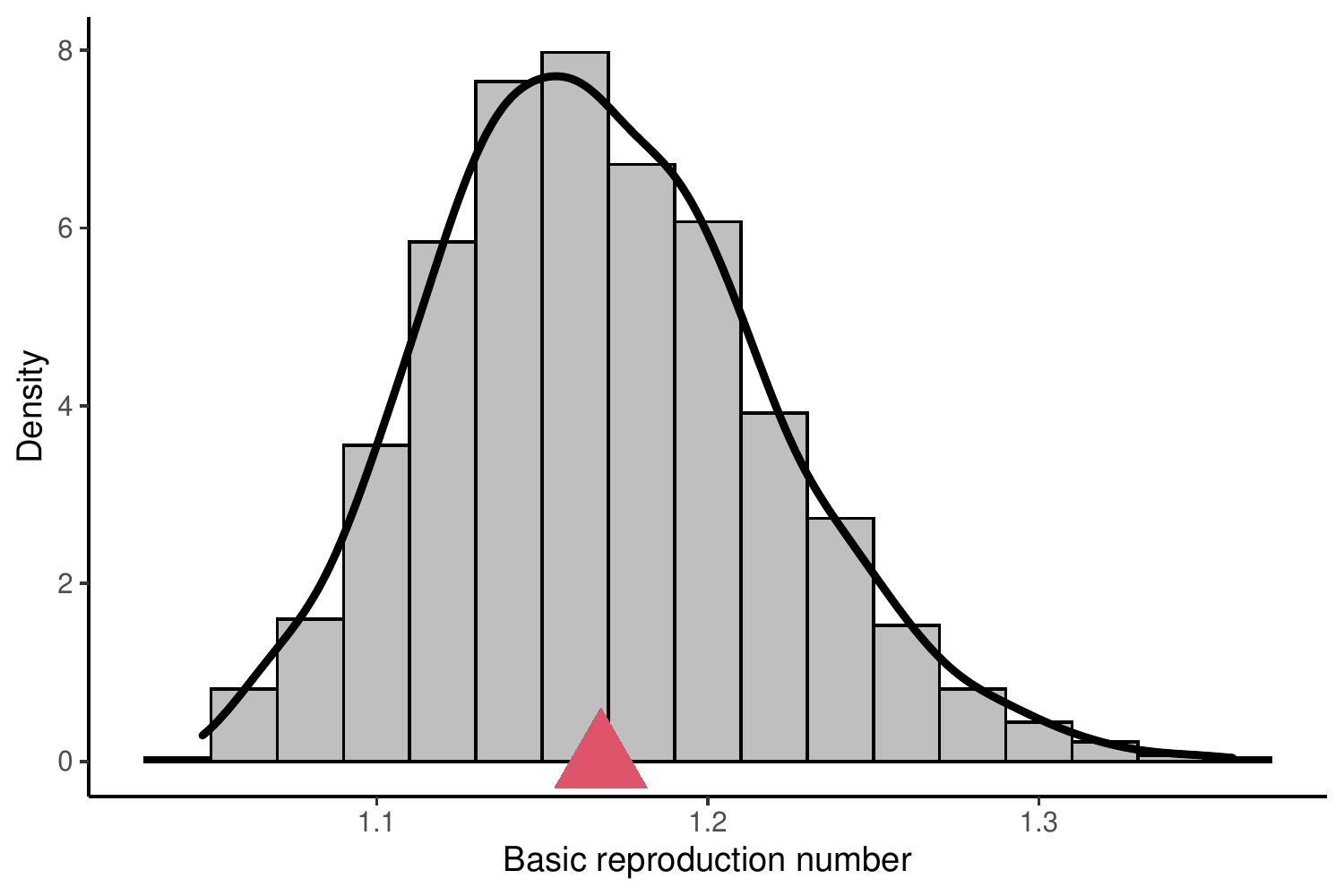}
    \includegraphics[scale=0.33]{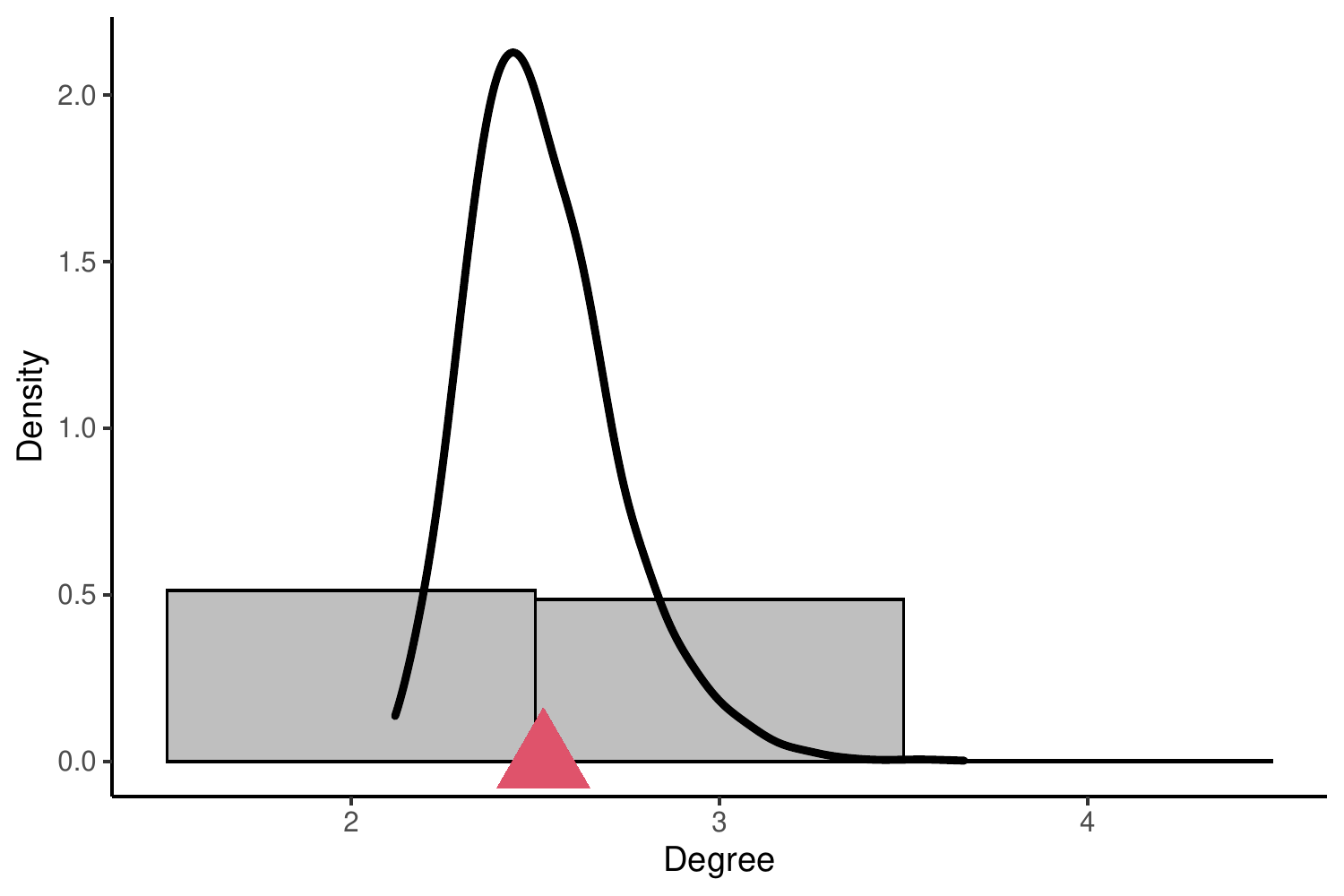}
    \includegraphics[scale=0.33]{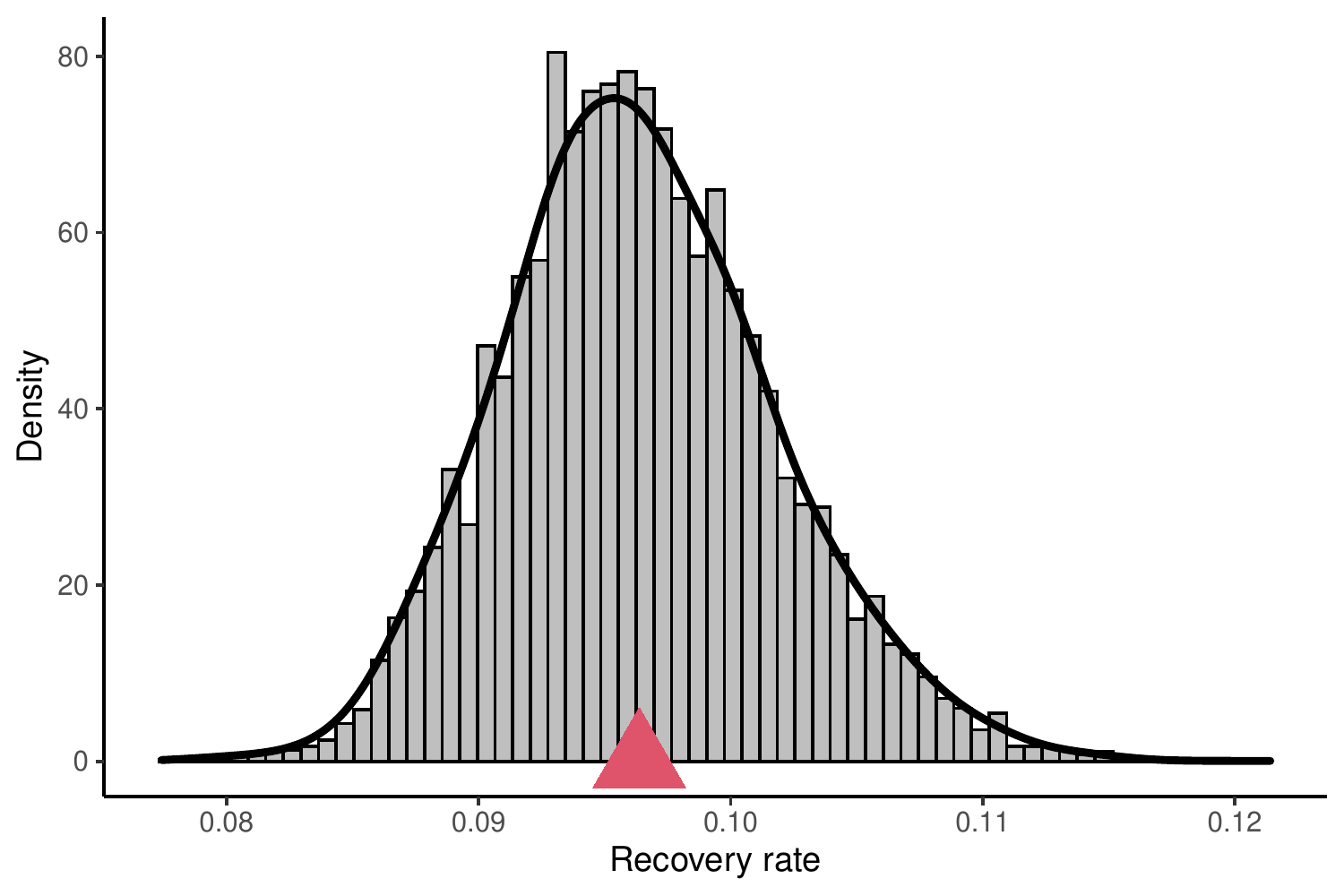}
    \includegraphics[scale=0.33]{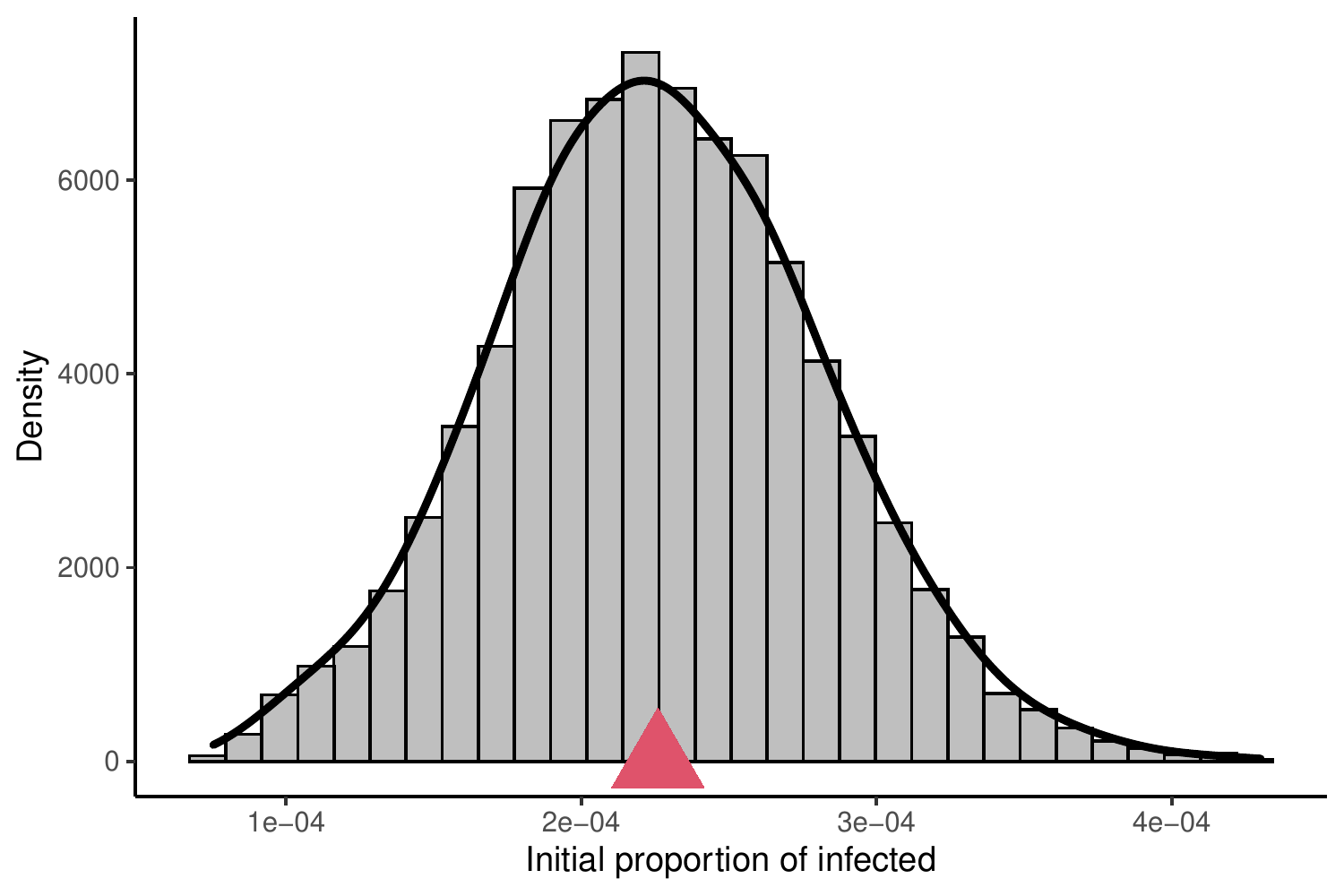}
    \includegraphics[scale=0.33]{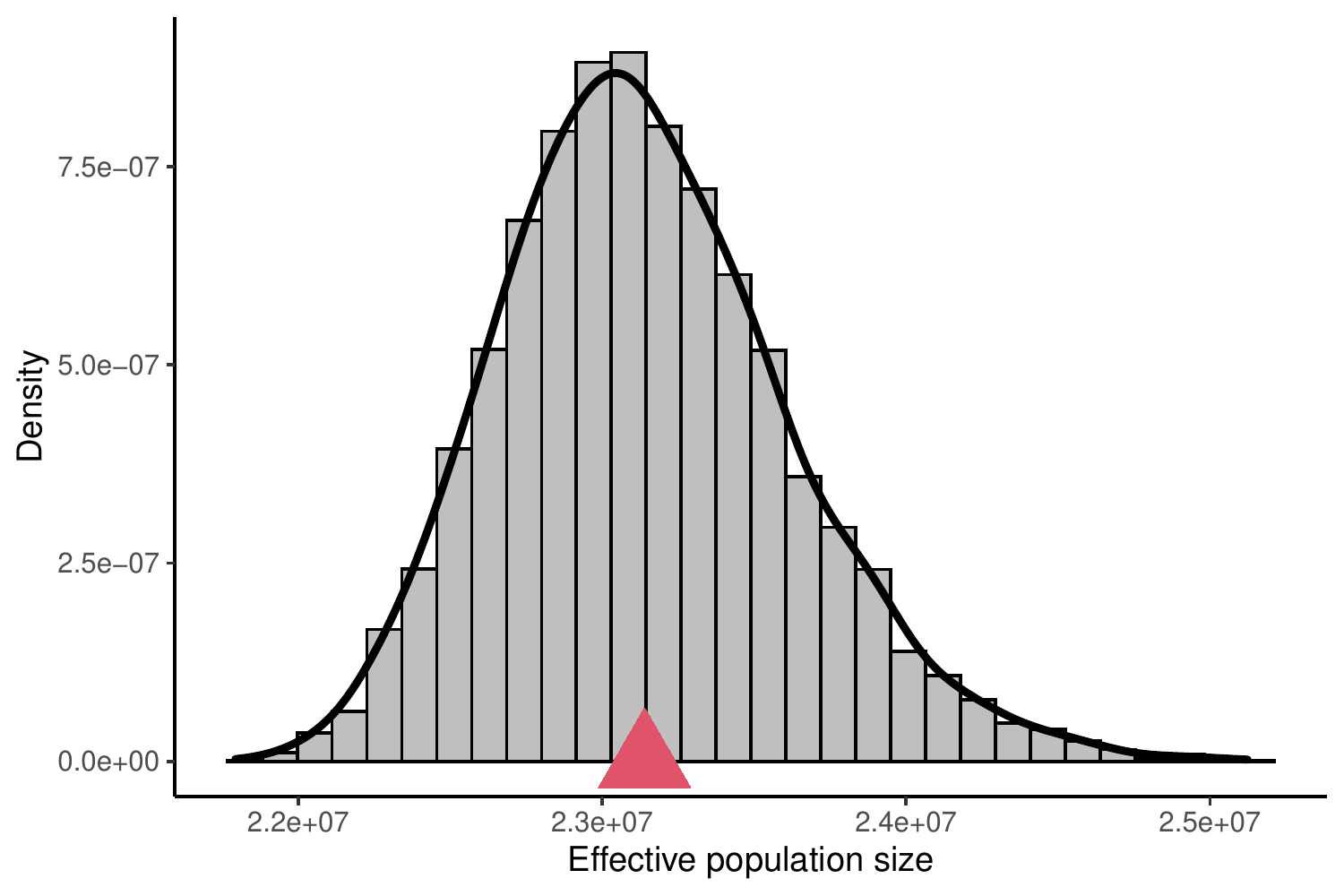}
    \caption{Posterior distributions of $(\tau, R_0, n, \gamma, \rho, n_T)$ using the DSA method on the COVID-19 dataset. The means and the medians of the posterior distributions are $(0.3745, 1.168, 2.522, 0.0963, 0.0002,   23139638)$ and $(0.3401, 1.164, 2.493, 0.0961, 0.0002, 23101076)$, respectively. }
    \label{fig:covid19_dsa_posteriors}
\end{figure}

As in the synthetic data study, random samples of individual infection  and recovery times (of size 5000 each) were constructed from the count dataset. These random samples were then fed into the HMC scheme using four parallel Markov chains. Uninformative, flat priors were used. 
The posterior distributions of the parameters $(\tau, R_0, n, \gamma, \rho, n_T)$ using the DSA method are shown in Figure~\ref{fig:covid19_dsa_posteriors}. The estimated parameters correspond to probability distributions that have similar measures of central tendency as those reported in an earlier analysis of the data in \cite{DiLauro2022NonMarkovDSA}. 

Interestingly, for both methods, the majority of the probability mass in the (posterior) distribution for the degree ($n$) is concentrated around small values, indicating a low contact pattern. This is in agreement with various non-pharmaceutical interventions such as lockdowns that were put in place to reduce the spread of the virus. Finally, both ML-estimated population size and DSA effective size are in the same order of magnitude. 

\subsubsection{Comparison across real-world datasets}
\label{subsec:MLEvsDSAcomparison}
\begin{figure}[h!]
	\center
	\includegraphics[scale=0.3]{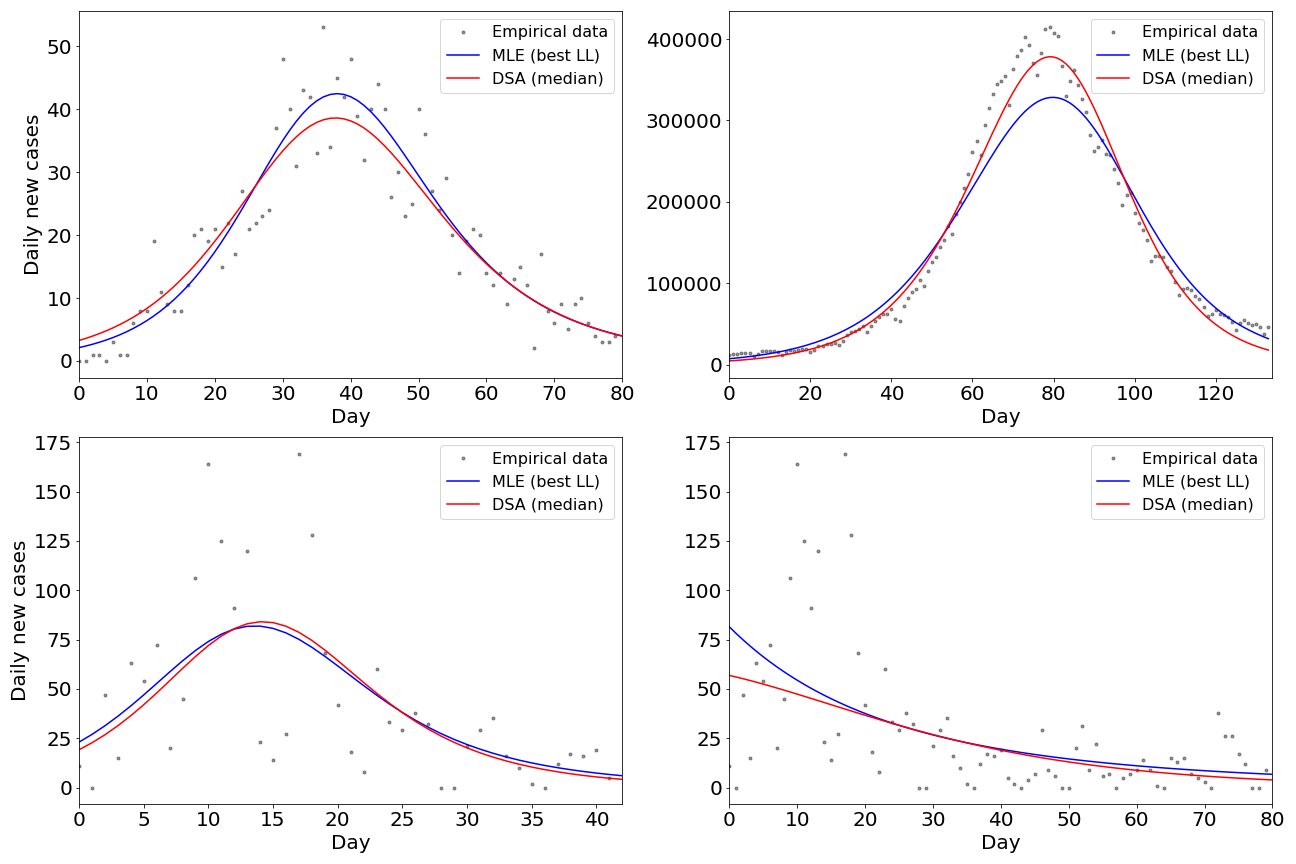}
	\caption{Illustration of the real-world outbreak data (top-left - 2001 FMD outbreak in the UK, top-right - third wave of COVID-19 in India, bottom panels - H1N1 outbreak with short (left) and long horizon (right)) together with output from the pairwise model with point estimates from MLE (values with best likelihood) and DSA (median values). All parameter values are given in Table~\ref{tab:empirical_MLE_stats}.}
	\label{fig:MLE_vs_DSA_fits}
\end{figure}

Figure~\ref{fig:MLE_vs_DSA_fits} shows the data for all three real-world outbreaks together with fits produced when taking the best parameter estimates using the ML-based approach and the median values of the posteriors produced by DSA. Whilst our investigation of the COVID-19 dataset supports a like-for-like comparison between inference schemes, there are differences in the way the analyses of the FMD and the A(H1N1) datasets were carried out. Specifically, whereas no prior was involved in the MLE-based approach, informative priors (based on published literature) were used for the Hamiltonian Monte Carlo scheme for DSA. This reflects an important and fundamental difference between MLE-based approach and DSA methodology (here implemented via a Hamiltonian Monte Carlo scheme), namely that the latter follows a Bayesian route. It should be noted, however, that the effect of the choice of priors should vanish in the limit of a large number of data points. 

With this in mind, we can make several observations:
\begin{itemize}
\item In general, the fit to the real data is good except in two cases. In the COVID-19 data, despite relatively similar parameters between methods, the DSA fit appears to capture the trend of the data a lot better than the MLE fit where a clear mismatch is being observed. The scenario in which the full H1N1 epidemic is subjected to inference highlights the challenge of highly variable, potentially noisy, data, as well as the impact of the observation period. In particular, as shown by the bottom two panels of Figure~\ref{fig:MLE_vs_DSA_fits}, the longer observation window allows the long and noisy tail of the epidemic to dominate, with both approaches missing the rise and fall in the daily new cases. 
\item Table~\ref{tab:empirical_MLE_stats} provides two sets of estimates for the MLE approach. As indicated previously, this is  for comparison purposes, the MLE process was repeated $100$ times using different initial conditions. The estimate denoted 'best' is therefore the 'true' MLE estimate (in the sense of being the one with maximum likelihood over all estimates of all rounds). Nevertheless, in the above, we kept all estimates provided their likelihood was close enough to that of the best one. In many cases, we observe a large difference between best and median. This is yet another manifestation of the unidentifiability problem whereby vastly different values of the mean-degree can result in likelihoods very close to the best one (i.e., with the same quality of fit). Interestingly, we note that, in general (a few estimates were excluded as per the text), the impact of unidentifiability did not affect $R_0$ as much as other parameters. 
\item The estimates for $I_0$ and population size, $N$, are relatively similar across both inference approaches, except for A(H1N1) when the full dataset is considered and COVID-19. For the A(H1N1) outbreak, the MLE method appears to overestimate $N$ by a large margin. Note that Washington State University campus is located in a relatively small town with a student population of size around 18000 and a resident population of size around 9000 \cite{KhudaBukhsh2020DSA}. For the COVID-19 wave in India, the DSA median estimate of 5204 for $I_0$ appears smaller than the true count of 11592 new cases on 16 February 2021, whereas the MLE method seems to overestimate it (median and best estimate of 33682 and 33130, respectively). It should be noted that the effective population size is a by-product of the DSA method (see Section~\ref{sec:DSA_principle}). Strictly speaking, the parameters $I_0$ and $N$ are far less meaningful in DSA than in MLE which requires them. However, keeping track of the DSA estimates $n_T$ of the effective population sizes at times $T$ is valuable in that it gives us a sense of the possible size of the epidemic and therefore, could be used for monitoring an ongoing epidemic \cite{khuda_bukhsh_2022_projecting}. 
\item Comparing the distributions obtained by DSA and MLE for the FMD data, we find the range of average degree obtained by DFA to be much better behaved than that obtained by MLE with mean and median being close and with a numerical value that seems more realistic. This observation holds for all datasets with DSA producing more realistic estimates. This is ultimately linked to the fundamental difference between how the likelihoods in the MLE and DSA approach are formulated. Whilst the MLE method simply minimises the mismatch between model trajectory and data, the DSA likelihood captures the underlying probability laws of individual infection and recovery times. More specifically, it models the underlying survival function through the $[S](t)$ curve parameterized by $(n, \tau, \gamma, \rho)$ (and implicitly, by the observation time $T$). 
\end{itemize}

\begin{table}[h!]
    \centering
    \begin{tabular}{l|r|r|r|r|r|r}
         & $I_0$ & $R_0$ & $n$ & $\gamma$ & $k$ & $N$ \\
         \hline
         \hline
        FMD (MLE median) & 11 & 2.71 & 26.32 & 0.0577 & 0.0123 & 1,748 \\
        FMD (MLE best) & 11 & 2.58 & 153.67 & 0.0723 & 0.0101 & 1,817 \\
        \hline
        FMD (DSA median) & 14 & 2.05 & 9.98 & 0.0737 & - & 1,819 \\ 
        \hline
        \hline
        H1N1-N18234 (MLE median, 42 days) & 76 & 2.69 & 229.81 & 0.1073 & 0.7688 & 2,098 \\
        H1N1-N18234 (MLE best, 42 days) & 76 & 2.70 & 2094.78 & 0.1073 & 0.7679 & 2,095 \\
        \hline
        H1N1-N18234 (DSA median, 42 days) & 39 & 1.85 & 10.91 & 0.1818 & - & 2,177 \\ 
        \hline
        \hline
        H1N1-N18234 (MLE median, 80 days) & 116243 & 2.07 & 1582.07 & 0.0142 & 1.2296 & 119,131 \\
        H1N1-N18234 (MLE best, 80 days) & 3463 & 0.85 & 2.61 & 0.0270 & 1.2468 & 20,256 \\
        \hline
        H1N1-N18234 (DSA median, 80 days) & 252 & 0.99 & 8.67 & 0.1818 & - & 9,286 \\ 
        \hline
        \hline
        covid (MLE median) & 33682 & 1.72 & 3.70 & 0.0323 & 0.0474 & 20,213,142 \\
        covid (MLE best) & 33130 & 1.70 & 3.68 & 0.0333 & 0.0474 & 20,254,332 \\
        \hline
        covid (DSA median) & 5204 & 1.16  & 2.50 & 0.0961 & - & 23,101,076 \\ 
        \hline
        \hline
    \end{tabular}
    \caption{Summary statistics of the inferred parameters for the three empirical datasets considered in this study when using both MLE and DSA approaches. Estimates for $I_0$ and $N$ were rounded to the nearest integer for readability.}
    \label{tab:empirical_MLE_stats}
\end{table}

\section{Discussion}
\label{sec:discussion}

In this paper, we have investigated the ability of a network-based mean-field model, i.e., the pairwise model, to infer not only disease parameters but also some of the underlying network. Outbreak data encapsulate the interplay between contact network and epidemic spreading. However, daily new cases or other data incorporate network information only implicitly. Hence, it is interesting to investigate whether from such data one can learn about the underlying contact network. Several challenges arise; for example, an epidemic with a small transmission rate on a dense network may look very similar to an epidemic with a large transmission rate spreading on a sparser network. Hence, it is not a given that outbreak data hold a specific enough signature of the contact network. In fact, our investigation revealed an anti correlation between the value of the transmission rate and the density of the network. Regardless, the estimate of both parameters peaked at around the desired values, especially when ground truth was known.

While the pairwise model used in the paper assumes that the network is regular and only accounts for the number of links each node has, it is possible to relax this seemingly restrictive assumption. In \cite{khuda_bukhsh_2022_projecting}, DSA was used for an SIR epidemic on a configuration model network with Poisson degree distribution. Recently, it has been shown~\cite{kiss2022necessary} that the pairwise model remains exact for networks with binomial, Poisson or negative binomial degree distribution; see also \cite[Corollary 1, Section 5.2]{khudabukhsh2022FCLT} where a similar result was derived for a susceptible-infected (SI) process on configuration model random graphs. The difference in the degree distributions manifests itself in the PW model via the type of closure one uses. For example, if the underlying network has a Poisson degree distribution, then  $\xi$ is simply set to $\xi=1$, and the parameter of the Poisson distribution, and hence, the network enters the PW model via the initial conditions. A similar modification is possible for networks where the degree distribution is negative binomial thus separating mean from variance. These all offer extensions and improvements above and beyond what the PW model was able to capture about the network. Moreover, employing the edge-based compartmental model, another network-based mean-field model, which uses the probability generating function of corresponding to the degree distribution of the network makes it possible to aim for learning the degree distribution of the underlying network.

The crucial advantage of the DSA methodology is the change in perspective about the mean-field ordinary differential equations. In the DSA approach, we view the ODEs as descriptions of probability laws of individual times of infection and recovery, as opposed to their traditional interpretations as limiting proportions or scaled sizes of compartments. By doing so, we are able to directly model the underlying survival functions corresponding to the individual times of infection and recovery, and thereby, bring to bear the entire toolkit of survival analysis for the purpose of parameter inference. Even though the DSA methodology has now been applied to several compartmental models, both Markovian and non-Markovian, both under mass-action and network-based contact patterns, the law of large numbers-based DSA methodology needs further improvement to adjust for stochastic effects when applied to finite (often small) populations. 

\section{Acknowledgements}
L. Berthouze and I.Z. Kiss  acknowledge support from the Leverhulme Trust for the Research Project Grant RPG-2017-370. The authors thank Prof Theodore Kypraios for useful discussions about approximate likelihoods.

\section{Data availability statement}
H1N1 outbreak data is available at \url{https://github.com/cbskust/SDS.Epidemic}, data about the third COVID19 wave in India can be found at \url{https://data.covid19india.org/documentation/csv/}. All other datasets generated during and/or analysed during the current study are available from the corresponding author on reasonable request.

\bibliography{project1}
\bibliographystyle{plain}

\end{document}